\documentclass[11pt,a4paper]{article}
\usepackage{jheppub} 
\usepackage{graphicx}
\usepackage{bm}

\usepackage{caption}
\usepackage{framed}
\usepackage{xcolor}
\usepackage{rotating}
\usepackage{empheq}
\usepackage{afterpage}

\usepackage[utf8]{inputenc}
\newcommand{\diff}{\mathrm{d}}

\newcommand{\Diff}{{\mathcal{D}}}

\def\Im{\mathrm{Im}}
\def\Re{\mathrm{Re}}

\def\C{{\cal C}}

\def\Z{{\mathbb Z}}

\def\mrmd{{\mathrm{d}}}
\def\mrms{{\mathrm{s}}}

\def\exp{\mathrm{exp}}

\def\mcalD{{\mathcal{D}}}
\def\mcalE{\mathcal{E}}

\def\V{{\cal V}}
\def\a{\alpha}

\def\g{\gamma}
\def\G{\Gamma}
\def\d{\delta}

\def\f{\phi}
\def\F{\Phi}

\def\n{\nu}

\def\r{\rho}

\def\t{\theta}
\def\x{\xi}

\def\tE{\tilde{E}}

\def\mbfr{{\mathbf{r}}}

\def\mbfB{{\mathbf{B}}}

\def\mbfq{{\mathbf{q}}}

\def\mcalM{{\mathcal{M}}} 
\def\mcalMbar{\overline{\mathcal{M}}}

\def\mbbZ{\mathbb{Z}}
\def\mbbR{\mathbb{R}}

\def\I{{\cal I}}

\def\*{\star}

\def\tA{\tilde A}
\def\tG{\tilde G}

\def\tr{{\mathrm{tr}}}

\def\<{\langle}
\def\>{\rangle}

\newcommand{\lbar}{\lower0.2ex\hbox{$\mathchar'26$}\mkern-10mu \lambda}

\title{\boldmath  	Polyakov Model in  't Hooft flux background: \\ 
 A quantum mechanical reduction   with  memory }

\author[a,b]{Cihan Pazarba{\c{s}}\i,}
\author[b]{Mithat \"Unsal}
\affiliation[a]{Physics Department, Bo\u{g}azi\c{c}i University, \\
 34342 Bebek, Istanbul, Turkey}
\affiliation[b]{Department of Physics, North Carolina State University, \\ Raleigh, NC 27695, USA}

\emailAdd{cihan.pazarbasi@gmail.com}
\emailAdd{unsal.mithat@gmail.com}

\abstract{We construct a   compactification of Polyakov model  on  $T^2 \times \mathbb R $  down to quantum mechanics which remembers non-perturbative  aspects of field theory even at an arbitrarily small area.  Standard compactification   on  small $T^2 \times \mathbb R $ possesses a unique   perturbative vacuum (zero magnetic flux state), separated parametrically from higher flux states, and  the instanton effects do not survive in the Born-Oppenheimer approximation.  
By turning on a  background magnetic  GNO  flux  in co-weight   lattice   corresponding to a non-zero 't Hooft flux, 
we show that  $N$-degenerate vacua appear at small  torus, and there are $N-1$ types of flux changing instantons between them.  
We  construct QM instantons starting with QFT instantons using the method of replicas.  
For example, $SU(2)$ gauge theory  with flux reduces to the double-well potential where each well is a  fractional flux state.  
Despite the absence of a  mixed anomaly, the vacuum structure of QFT and the one of QM are continuously connected.    We also compare the quantum mechanical reduction of the Polyakov model  with the deformed Yang-Mills, by coupling both theories to 
TQFTs.   In particular, we compare the mass spectrum for dual photons and energy spectrum in the QM limit.  We give a detailed description 
of critical points at infinity in the semi-classical expansion, and their role in resurgence structure. 
}

\begin{document}
    \maketitle
    \flushbottom
    
    \section{Introduction} 
    Polyakov model is  a well-known  non-abelian gauge theory on $\mathbb R^3$ with  an  algebra valued Higgs scalar in the adjoint representation. The model,  studied by Polyakov in mid-70s,  is the first example of calculable confinement and mass gap  in a $d \geq 3$ gauge field theory \cite{Polyakov:1976fu}. The non-perturbative phenomena  can be described by using semiclassical methods, in terms of   proliferation of monopole configurations  (instantons in 3d, not particles),  which are   leading order saddles  in semi-classics in Euclidean path integral formulation. 
    Even though this story  is fairly well-known, and a textbook material by now \cite{Deligne:1999qp,  Banks:2008tpa, Tong},  some important aspects of  semi-classical analysis remain to be unsatisfactory. Especially the role  of critical points at infinity (and relatedly correlated instanton events)  are usually pushed under the rug.    For example, third-order in semi-classics produce an imaginary ambiguous contribution to the mass gap  
$\Delta m_g^2=c_1 e^{-3 S_0} \pm i \pi    e^{-3 S_0} $  that is not even mentioned in old literature,  
 and  would be disastrous for Polyakov's analysis as it stands  \cite{Pazarbasi:2021ifb}.

\begin{figure}[t]
    	\begin{center}
    		\includegraphics[scale=0.4]{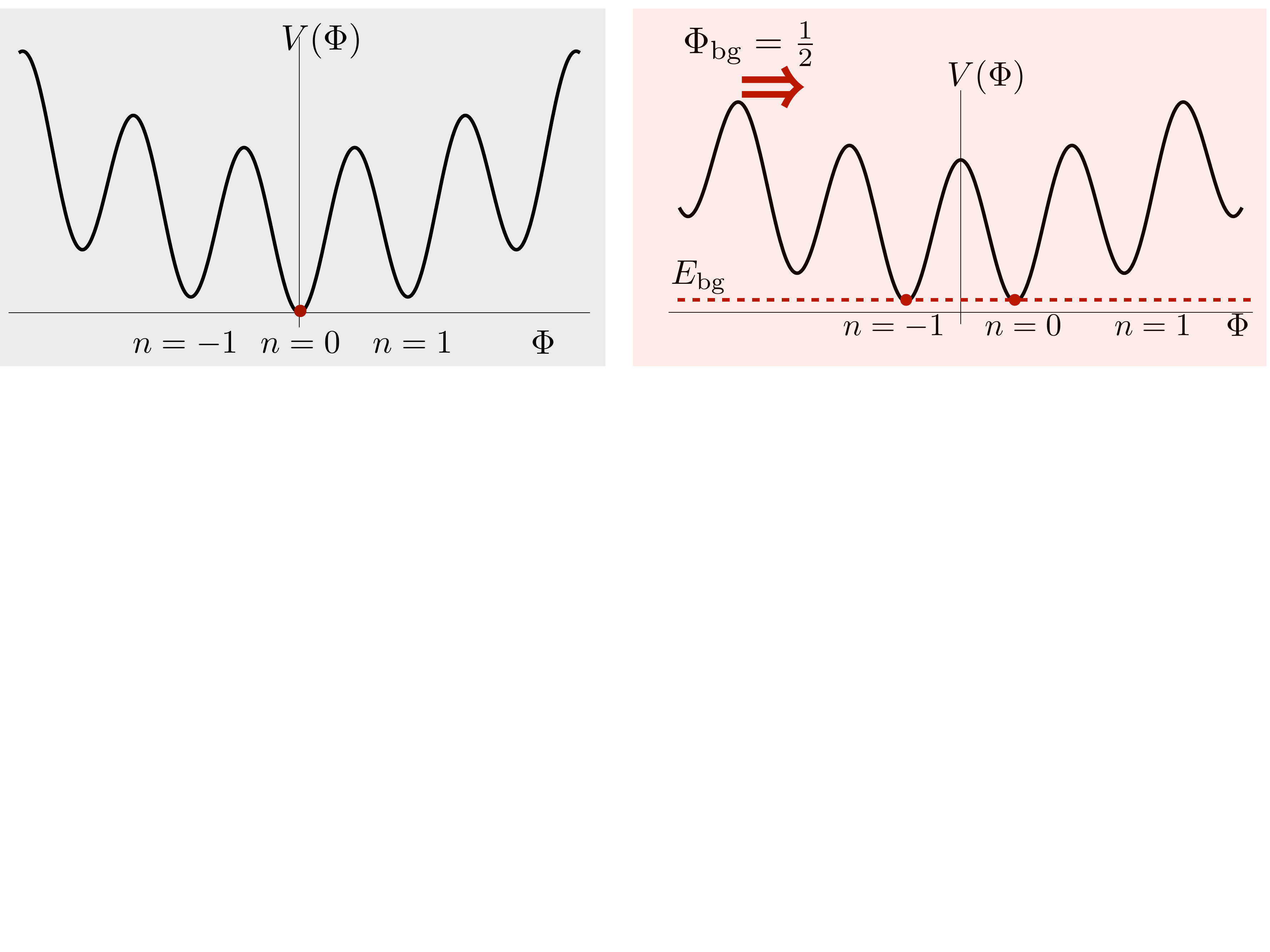}
    	\end{center}
    	\vspace{-6.3cm}
    	\caption{Compactification of  $ SU(2)$  Polyakov model to QM  for $\Phi_{\rm bg} =0$ and  $\Phi_{\rm bg} =\frac{1}{2}$ background magnetic GNO flux. $\Phi_{\rm bg} =\frac{1}{2}$ GNO flux corresponds to one-unit of 't Hooft flux. 
	 In the former, 
    		the vacuum is unique perturbatively at finite $A_{T^2}$, and degeneracy only emerges in the $A_{T^2}  \rightarrow \infty $ limit. 
    		In the latter case, there  is a two-fold degenerate vacua perturbatively, and  instantons persists  even at small  $A_{T^2} $.  Generalizing this 
    		structure  to  $SU(N)$ theory, we can achieve an $N$-fold perturbative degeneracy in the QM limit, and $N-1$ types of instantons.  }
    	\label{wowy}
    \end{figure}

Another  interesting question  in the Polyakov model that is partially  addressed in textbooks  (see Banks' book or Tong's lecture notes  \cite{Banks:2008tpa, Tong}) is following.  What classical configurations the monopoles  are tunneling in between? 
  	Compactifying the theory on $T^2 \times \mathbb R$,   the flux through $T^2$,   $\Phi= \frac{2 \pi}{g} n,  \; n \in \mathbb Z$ is quantized,  (these are  magnetic flux of monopoles  valued in co-root lattice of $SU(2)$) and  the configurations with non-zero flux  have higher energy than the ground state:  
 	\begin{equation}\label{Energy_Mon-zerofinite}
    		\Delta E_{n, 0}  = \frac{1}{2}\int_{T^2} \bm B^2 = \frac{1}{2 A_{T^2}}    \left( \frac{2 \pi}{g} n\right)^2  > 0
    	\end{equation}	
	where $A_{T^2}$ is the area of torus.  	See Fig.~\ref{wowy}, left.     Therefore,  zero flux vacuum and higher flux configurations   are non-degenerate for finite values of $A_{T^2}$.   Refs.\cite{Banks:2008tpa, Tong} point out that these flux states  become degenerate in the   large $T^2$ limit and the monopoles are the instantons which tunnel between these states. A consequence of the analysis of  Refs.\cite{Banks:2008tpa, Tong} is that 
    	in the quantum mechanical small $T^2$ limit, the ground state is  unique in perturbation theory, and within the Born-Oppenheimer approximation, the low energy limit  is described by a single-well potential with no memory of instantons, corresponding to central well in Fig.~\ref{wowy}.
	   In this paper, we pose the following question: 
	\begin{quote}
	{Is there a compactification of the $SU(N)$  Polyakov model in which $N-1$ types of   instantons on $\mathbb R^3$  and  corresponding perturbatively degenerate vacua survive at finite 
and  small  $T^2 \times \mathbb R $ in a  quantum mechanical limit? 	}
\end{quote}
Remarkably, the answer to this question turns out to be  positive. And the construction goes through another   time-honored method  from mid 70s, the use of  
 background  't Hooft  fluxes \cite{tHooft:1979rtg, tHooft:1977nqb,vanBaal:1982ag}.   
 This method is applied successfully  to understand  Yang-Mills dynamics in  lattice gauge theory to exhibit fractional instantons  \cite{ GarciaPerez:1992fj, GarciaPerez:1993jw, Gonzalez-Arroyo:2019wpu}. 
  It also plays an important role in large-$N$ volume independence via TEK models \cite{GonzalezArroyo:1982hz}, which, in retrospect,  provides  a non-perturbative lattice formulation  of non-commutative field theory \cite{Ambjorn:2000cs}.     't Hooft  fluxes also  enter to recent literature as a  particular realization of 
 coupling a topological quantum field theory (TQFT) to QFT \cite{Kapustin:2014gua}.    
Up to our  knowledge,  and surprisingly,  
Polyakov model is not examined in  't Hooft flux background since their more or less simultaneous  inceptions about four decades ago.  In this work, we explore  benefits of merging these two ideas. 

%
    	
 The way we implement 't Hooft flux is slightly different from  't Hooft original construction and other works.    Our construction  takes advantage of the dynamical abelianization of the Polyakov model \cite{Unsal:2020yeh}.   
 In particular,    we turn on a background magnetic GNO flux  \cite{Goddard:1976qe}, e.g. for $SU(2)$    $\Phi_{\rm bg}= \frac{1}{2}$ through $T^2$, which lives in co-weight lattice.  Since co-weight modulo co-root  $ \Gamma_w^\vee/  \Gamma_r^\vee \cong \Z_N$,  for $SU(2)$  this is equivalent to the insertion of one unit of discrete  't Hooft flux.  And indeed,  something   interesting  happens.  The energy of configurations are modified into 
 \begin{align}
   E_n = 
    \frac{1}{2 A_{T^2}}    \left( \frac{2 \pi}{g} \Big( n +\frac{1}{2} \Big) \right)^2, \qquad \Delta E_{n=-1, 0} =0.
 \end{align}  
   The   $n=-1$  and $n=0$ states become degenerate   classically.  
     In particular, in the Born-Oppenheimer approximation, the system is described by a double-well potential, and monopole instantons survive in the compactification with 't Hooft flux even at small  
    	$A_{T^2}$ limit.  See Fig.~\ref{wowy}, right.  Insertion of 't Hooft flux  is useful for  the remembrance of the things that exist in QFT in the 
	quantum mechanical compactification or reduction. 
    	
 One importance of our  quantum mechanical reduction (which remembers the instantons and vacuum structure of the infinite 
    volume theory)  is in connection to resurgence properties \cite{Dunne:2016nmc}. 	  In particular, 
    the correlated  $[\mcalM \,\mcalMbar]_{\pm}\; $ configuration (which is part of the  thimble integration of the  critical point at infinity  as we make precise) is  two-fold ambiguous \cite{Bogomolny:1980ur,  Zinn-Justin:1981qzi, Dunne:2014bca, Behtash:2018voa} in QFT. It is difficult to demonstrate the cancellation of this ambiguity in full QFT because the large-order perturbative data and its ambiguity upon Borel resummation are not available.  However, by building a QM reduction 
   (which possess the same action instanton configurations),   we can show  all orders resurgent cancellations in multiple ways, most efficiently 
  by using the exact WKB method   \cite{DDP, Sueishi:2021xti}. Even though this does not solve the problem in QFT, it  makes it 
   natural to conjecture that   the same resurgent cancellation is also present in full QFT.  Based on this conjecture, we provide  estimates  of large orders of perturbation theory in QFT. 
    	
    	
 Mass gap and string tension in the Polyakov model are induced by proliferation of the $\mcalM$  and $\mcalMbar$ events, e.g.  $m_g^2 \sim  e^{- S_0} $.    However, there are certain  critical points at infinity,  and their thimbles 
  include  correlated     $[\mcalM\mcalM\mcalMbar]_{\pm}$  and    $[\mcalM\mcalMbar\,\mcalMbar]_{\pm}$  effects,  
	which carry the same magnetic charge as $\mcalM$  and $\mcalMbar$, respectively.  One may think that these are $e^{- 3S_0} $ effects and can be   neglected as argued  in Coleman's lectures \cite{Coleman198802}.\footnote{
	As a matter of fact, the generic confinement mechanism in gauge theories on $\mathbb R^3 \times S^1 $ is due to second order   
	($e^{- 2S_0}$)
	effects in semi-classics, called magnetic bions. Monopole-instantons, which appears at first order 	($e^{- S_0}$),   do not typically induce confinement due to their fermion zero mode structure. See \cite{Shifman:2008ja}.}
    	However, the non-trivial issue is that these configurations give a {\it qualitatively} new effect; they are   multi-fold ambiguous, and their contribution  	must be included.   	  Their contribution to mass gap is ambiguous, of the form   $\Delta m_g^2=c_1 e^{-3 S_0} \pm i \pi    e^{-3 S_0} $. As it stands,  this would  be  an important problem concerning the  mass gap in the Polyakov model.  
  We are supposed to obtain a real mass gap, but we obtain corrections that have imaginary ambiguous parts!  This is not something that one can be dismissive about.  In  our QM reduction, we show the resurgent cancellation of such 
  ambiguities, leading to meaningful results.  
    \noindent
    	   \begin{figure}[t]
    		\vspace{-1.3cm}
    	\begin{center}
    		\includegraphics[scale=0.35]{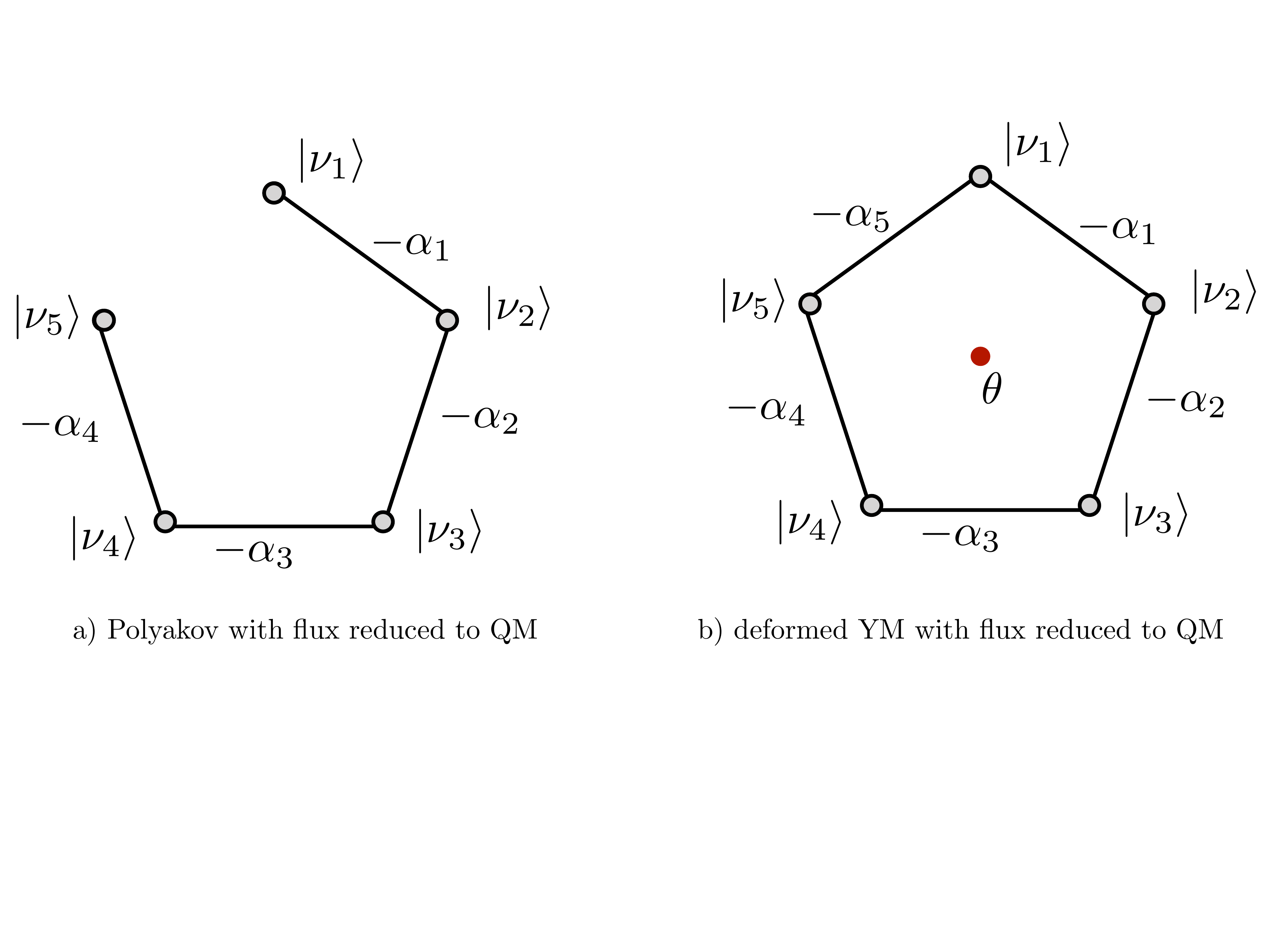}
    	\end{center}
    	\vspace{-3.4cm}
    	\caption{ Dimensional reduction of 
	$SU(N )$   Polyakov   and dYM with the insertion of 't Hooft flux down to quantum mechanics on $T^2 \times \mathbb R$ and  $T^2 \times \mathbb R \times S^1$, respectively.  $N=5$ above.
	Both theories  have $N$-perturbative vacua,   $N-1$   ($N$)  monopoles  are  associated with the simple  (affine) 
	 root systems. 
	 In the absence of  't Hooft flux,  there is a unique perturbative vacuum and monopoles  of the QFT do  not survive in QM limit in Born-Oppenheimer approximation.
	 }
    	\label{PDYM}
    \end{figure}
    
    These properties  generalize to $SU(N)$ gauge theory Higgsed to $U(1)^{N-1}$. In this case,  with the insertion of a non-trivial  background magnetic flux in co-weight  lattice $\Gamma_{w}^{\vee}$,  $N$ distinct perturbative flux  vacua remain degenerate in perturbation theory, and there are $N-1$ instantons connecting them as shown in Fig. \ref{PDYM}.  Therefore,  't Hooft flux allows the instantons in QFT to survive in the QM limit even  on small $T^2$.\footnote{In large volume, turning on a discrete 't Hooft flux background  does not alter local dynamics. But on small space, turning on  
 an appropriate    't Hooft flux changes  dynamics in a better  way relative  to its absence!   It helps   the topological excitations of the  large-volume theory to survive in a small volume limit.  In this sense, perhaps, it is better  to perform compactifications with 't Hooft flux insertion  in general. This is also the perspective of  old lattice works  \cite{ GarciaPerez:1992fj, GarciaPerez:1993jw, Gonzalez-Arroyo:2019wpu}.}
     This idea is  similar to the role that global symmetry twisted boundary conditions play in sigma models on $\mathbb R \times S^1$  
     \cite{Dunne:2012ae,  Dunne:2012zk,  Shifman:2014fra,   Fujimori:2017oab, Fujimori:2018kqp,   Fujimori:2016ljw, Misumi:2016fno, Misumi:2014jua, Misumi:2019upg, Krichever:2020tgp} and the role that center-symmetric holonomy plays on 
     $\mathbb R^3 \times S^1 $, see e.g. \cite{Dunne:2016nmc, Poppitz:2021cxe}.

    This work can be viewed as  an extension of   the  recent study in  $SU(N)$  Yang-Mills theories in 
    $\mbbZ_N$ TQFT background formulated 
    in $\mbbR\times S^1 \times T^2$  \cite{Unsal:2020yeh}  with center-stabilizing double-trace deformation in the $S^1$ circle.  
    See also \cite{Yamazaki:2017ulc} for related ideas that connect YM to  sigma model with global symmetry twisted boundary conditions     \cite{Dunne:2012ae}. 
    The deformation on $S^1$ circle ensures adiabatic continuity between small   $\mathbb R^3 \times S^1$  and $\mathbb R^4$ \cite{Unsal:2008ch}.  This is already checked on the lattice and it works, see e.g. \cite{Bonati:2018rfg}.   In both dYM, and Polyakov, the long-distance dynamics abelianize to $U(1)^{N-1}$. In this sense, we take advantage of two aspects: 
    \begin{itemize}
  \item  [ {\it 1)}]
   Dynamical abelianization and semi-classical calculability;   
   \item [{\it 2)}]
Background magnetic flux  valued in  the co-weight lattice $\Gamma_w^\vee$  and  corresponding discrete 't Hooft flux, which are related because 
$  \Gamma_w^{\vee} / \Gamma_r^{\vee}  \cong   \Z_N  $.  
\end{itemize}

    There are multiple  differences between the present work and Ref.   \cite{Unsal:2020yeh}. In the latter,   the Higgs scalar is group valued (instead of being algebra valued),  and there is also   a topological $\theta$  angle.  As a result, within Born-Oppenheimer approximation, the $SU(2)$ Polyakov model reduces to double-well potential on $\mathbb R$, and 
    deformed YM on $\mbbR\times S^1 \times T^2$ reduces to  
  the particle on a circle $S^1$  with two-minima on the circle.    The relation between these two constructions is discussed  in Section.~\ref{compare} and depicted in Fig. \ref{PDYM}. 
    \section{Quick summary of  $SU(N)$ Polyakov Model}\label{Section: PolyakovModel}
    We first briefly review the $SU(N)$ Polyakov model \cite{Polyakov:1976fu}, and  set the notation for the rest of the paper. 
  Consider a  $SU(N)$ non-abelian gauge theory coupled to an adjoint scalar field with   Euclidean action:
    \begin{equation}\label{PolyakovModel_Action}
    	S = \int \frac{1}{2 g_3^2}\bigg(\tr \left( F \wedge \*F \right) + \tr\left( D \f \wedge \* D \f\right)    +V(\f)  \bigg) 
    \end{equation}
    where \[F =D A \quad , \quad D = \mrmd + i A . \]    
    We choose the vacuum structure of $V(\f)$ such that the vacuum expectation value of the adjoint valued Higgs scalar is given by  
    \begin{align}\label{PerturbativeVacuum}
    	\f_v =\begin{bmatrix}
    		v_1 &  & & \\
    		& v_2 & & \\
    		&  & \ddots  & \\
    		&  &  & v_N  \\
    	\end{bmatrix}
    	= \begin{bmatrix}
    		-\frac{N-1}{2} v &  & & \\
    		&   -\frac{N-3}{2}v & & \\
    		&  & \ddots    & \\
    		&  &  &  + \frac{N-1}{2}v \\
    	\end{bmatrix}  
    \end{align} 
    %
    and  leads to the breaking of $SU(N)$ symmetry to $U(1)^{N-1}$. 
    For example, for $SU(2)$ theory,  we take $ V(\f)  =  \lambda  (\tr (\f^2)  - \frac{1}{4}v^2 )^2 $ where $\lambda$ is dimensionless. We chose the distance between the eigenvalues to be $v$, but this can be relaxed through the discussion as long as  $v_i$ are non-degenerate. 
     For convenience, the classical mass dimensions of fields and parameters in the theory are: 
    \begin{align}
    	{\rm mass \; dimensions:}  \qquad [g^2_3]= [A] = [\f] =+1,  \;\;\;  [\lambda]=0
    \end{align}
    As a result of the adjoint Higgsing, the perturbative spectrum contains $N(N-1)$ massive gauge bosons, or $W$-bosons,   and    $N-1$ massless gauge bosons corresponding to the photons of the unbroken $U(1)^{N-1}$ gauge structure.  For simplicity of the discussion, and without loss of generality,  we will also take   $|v_i - v_{i+1}| = v $. As a result, the masses of the  $ N-1$ lightest $W$-boson masses are equal to each other and given by 
   \begin{equation}
    	m_{W} 
    	= |v^i -v^{i+1}|   = v
    \end{equation}
    There are also $N^2-1$ massive scalars. These scalars also acquire masses due to adjoint Higgsing.   The lightest      $(N-1)$ scalar masses are given by     
    \begin{align}
        m_\f \sim  \sqrt \lambda  v \qquad  (\rm lightest \;\; scalar \;\; mass) 
        \end{align}
         The  lightest scalars may be heavier or lighter than the lightest $W$-bosons depending on the value of $\lambda$. 

\vspace{0.3cm} 
  The theory, at low energies can be described as a $U(1)^{N-1}$ abelian gauge theory to all orders in perturbation theory. However, non-perturbatively, 
  the gaplessness is destabilized by the monopole-instantons, the leading non-perturbative saddles in the problem. 
  Constructing a dilute gas of these instantons and anti-instantons, Polyakov \cite{Polyakov:1976fu} showed that the dual photons acquire a non-perturbative  exponentially small mass.  Below, we briefly describe this phenomenon, and then, we will move to a more subtle subject of critical points at infinity. 
  
%
%
\vspace{0.3cm} 
In the vicinity of  BPS limit, where $\lambda \ll 1$,  \cite{Weinberg:1979zt} monopole-instantons are solution to the self-duality equations: 
\begin{equation}
	F  = \pm \*_3 D\f .
\end{equation}
The minimal action monopoles are associated with the simple root system. They carry a magnetic charge 
\begin{equation}\label{charge}
	Q_{M_i} =
	 \frac{2\pi}{g_3}\a_i 
\end{equation}	
and the classical actions of the $\alpha_i$ monopole takes the values
\begin{equation}\label{Action_minimum}
  	S^{(i)}_0
	=   \frac{4 \pi v}{g_3^2}  \equiv   \frac{\mrms_0} {g^2}
\end{equation}
where we defined the dimensionless coupling $g^2 = \frac{g_3^2}{v}$ and the constant $\mrms_0 = 4\pi$. For later purposes, 
describing monopole interactions, it is also useful to define dimensionless position $\tilde r = v r$. 

Away from the BPS limit,  where $ \lambda$  can take any value, the coupled 
second order Euclidean equations of motions, $\frac{\delta S}{\delta A_{\mu}}=0, \; \frac{\delta S}{\delta \f}=0$
 also admit  solutions associated with magnetic charge  \eqref{charge}.  In that case, the action of the monopoles are given by
    \begin{align}\label{action-lambda}
    	S_0(g^2, \lambda) = \frac{4\pi }{g^2} f(\lambda), \qquad {\rm where}  \;\;\;  f(0)=1, \;\;\;  f(\infty)=1.787
    \end{align}  
    where $f(\lambda) = 1+ \frac{ \sqrt \lambda}{2} + \ldots $ \cite{Kirkman:1981ck} at small  $\lambda.$   Even at arbitrarily large-$\lambda$, the action of  the elementary monopole saddles  does not  exceed  $1.787 S_0 $.  
    
    \vspace{0.3cm} 
    The monopole core size  is the inverse $W$-boson mass, $ r_m \sim m_W^{-1}  =v^{-1}$.  At distances larger than $r_m$, the dynamics should be described     by an abelian gauge theory.  
    In 3d, one can use abelian duality to express the $N-1$ photons in terms of dual scalar  $\bm \sigma $. The reason for using dual fields is that the long distance  EFT and the  monopole operators has a natural representation  in terms of dual photons.  The abelian duality relation is  $ \star_3{ \bm F}=  \frac{g^2}{2 \pi}   \mrmd { \bm \sigma} $ where ${ \bm F}$ denotes Cartan components of non-abelian gauge field.  In the $SU(N)$ theory, the    $\bm{\sigma}$ field   has  a periodicity  determined by weight lattice,   $\bm{\sigma}\sim \bm{\sigma}+2\pi \bm{\mu}_i ,  \; \bm{\mu}_i  \in \Gamma_w$, and 

	\begin{align}
	\bm \sigma  \in \frac{\mathbb R^{N-1}}{2 \pi \Gamma_w}
	\label{s-cell}
	\end{align}
is the fundamental domain of $ \bm \sigma $.   

\vspace{0.3cm} 
	There are    $N-1$ types of the fundamental monopole instantons that appears as leading  order saddles, and  the monopole operators are given by\footnote{ If the diagonal of the adjoint scalar remains light or gapless for some reason (e.g. as  in ${\cal N}=2$  SYM  theory on $\mathbb R^3$ \cite{Affleck:1982as}), it should also be included in EFT and the monopole operator should be modified into  $e^{ - \frac{4 \pi}{g^2} \tilde {\bm \phi} (x) \cdot {\bm \alpha}_i + i \bm{\alpha}_i \cdot \bm{\sigma}(x)}$. In the present case,   	it is consistent to set the fluctuations  $  \tilde {\bm \phi} (x)  $ to zero once we consider the physics at  long distances. }
    \begin{align}
    	{\cal M}_{\a_i}(x)= K e^{-S_0}   e^{i \bm{\alpha}_i \cdot \bm{\sigma}(x)} \equiv \xi  \;  e^{i \bm{\alpha}_i \cdot \bm{\sigma}(x)}   \qquad i=1, \ldots, N-1 .
    	\label{mon-op}
    \end{align}       
%
    Since monopoles are finite action, they proliferate in the Euclidean vacuum. The density  of the monopoles of any type is given by $\xi \sim e^{-S_0}$.  
    The grand canonical ensemble of the monopole gives the partition function: 
    \begin{align}
    	Z= \int D \bm \sigma  \;  \exp  \Big[  -\Big(   \int {g^2\over 8\pi^2 } | d \bm{\sigma} |^2 -
    	2 \sum_{i=1}^{N-1}  \xi \,  \cos\left(\bm{\alpha}_i  \cdot \bm{\sigma} \right) \Big) \Big]
    	\label{master}
    \end{align}  
    This partition function should be viewed with a UV cut-off which is of  the order of inverse  monopole core size, 
    \begin{align}
    	{\rm UV \; cutoff \; in \; monopole \;  EFT:} \qquad   \Lambda_{\rm UV} \sim  r_m^{-1} \sim  m_W .
    \end{align}
    The EFT  \eqref{master}  based on monopoles generates mass gap for $N-1$ dual photons. The masses  of $N-1$ types of dual photon can be obtained by diagonalizing quadratic part of the potential and is given by
     \begin{align} 
  m_k^2  = m_\gamma^2  \sin^2  \left(   \frac{ \pi }{2N} k  \right),    
  \qquad k=1, 2, \ldots, N-1 
 \end{align}
where $m_\gamma^2 \sim   \xi / g^2$. The fact that the mass gap is sourced by the fugacity of the monopole-instantons is the well-known result of Polyakov.

    
\section{Critical points at infinity and cluster expansion}\label{Section: CriticalPointsAtInfinty}
    In the Polyakov model, the instantons with a magnetic charge, as described above, are solutions to either BPS equations (as $\lambda \rightarrow 0$) or full second-order coupled equations of motions. Their proliferation leads to a dilute gas of monopoles described in \eqref{master}, which is first order in $e^{-S_0}$.  However, as in {\it cluster expansion} in statistical mechanics \cite{kardar}, some effects arise due to correlated events or molecules of instantons.  These are higher-order effects in semi-classics. We briefly review the cluster expansion in Appendix \ref{Section: ClusterExpansion}. 
    
    \vspace{0.3cm} 
    Even though the  mass gap is  of order $m_g^2 \sim e^{-S_0}$ at leading order in semi-classics,  there are qualitatively new effects at higher order.  For example, at third-order semi-classics, there are contributions to the mass gap that are of the form  $\Delta m_g^2 \sim e^{-3S_0	}  \pm i \pi e^{-3S_0}$, where the imaginary part is    two-fold ambiguous.  These phenomena usually  manifest themselves in a careful treatment of critical points at infinity.\footnote{The definition of the critical point at infinity is as follows: Consider a  two instanton configuration with a (repulsive or attractive) interaction in between.   At any finite separation,  the combination 
     is not a solution, because equations are non-linear and superposition of the solutions is not a solution, i.e, $\frac{d  V_{\rm int}(r)}{ dr } \neq 0$ at any finite separation. 
  But  it becomes a genuine saddle at infinite separation because   $\frac{d  V_{\rm int}(r)}{ dr } |_{r= \infty} =0 $. If we map $r \in \mathbb C$ complex domain to Riemann sphere by using one-point compactification,  the critical point is the North pole and is on the same footing with any other regular critical point in a certain sense.  The feature that makes this  saddle special is its  non-Gaussian nature. Because of this property, one needs to integrate over the whole steepest descent cycle to obtain qualitatively correct  non-perturbative contributions. Almost all saddles in QFT and QM are critical points at infinity.  Yet, they are the least appreciated and understood ones. See Ref.~\cite{Behtash:2018voa} in the context of QM. }

\vspace{0.3cm} 
    In this section, we will describe some features of critical points at infinity and  correlated $k$-events.  In particular, we will assert that some of these ambiguous correlated events are related to the resurgence properties of QFT and cancel in the full resurgent semi-classical analysis. However, it is  hard to prove this statement in  full QFT. Yet in many  quantum mechanical examples, our understanding of all order resurgent cancellations is much better.  Therefore, it is desirable to reduce QFT  to QM while keeping its non-perturbative aspects as intact as possible.  

\vspace{0.3cm} 
Up to our knowledge. no such compactification of the Polyakov model is known so far.   The regular compactification  with periodic boundary conditions  on {\it small} 
    $T^2 \times \mathbb R$ has a single perturbative vacuum and no memory of the  instanton events in the Born-Oppenheimer approximation.\footnote{However, if we consider flux $ |\Phi= \frac{2 \pi}{g} n \rangle,  \; n=\pm 1$ states in QM limit, these are degenerate at finite box too.  The tunneling between them is  
    two-monopole events with $\Delta n=2$,  see  Fig.~\ref{wowy}. But these states do not survive in BO approximation, and in the description of ground-state properties of QM.}    
      In the next section, to circumvent this problem, we will construct the Polyakov model in 't Hooft   flux background on small 
    $T^2 \times \mathbb R$ that remembers the instanton events of full field theory.  In that context, we can prove for example  all order resurgent properties by using exact-WKB formalism. This 
    leads us to a conjecture that the same resurgent cancellations also take place in the context of full QFT. Assuming this is the case, we will provide predictions of the large-order growth around the perturbative vacuum and  monopole-instanton saddles. 
	
    \paragraph*{Correlated and uncorrelated 2 instanton events:}      How to describe the critical point at infinity and correlated instanton events? Let us start with the  possible combinations at  the  two instanton order, eg.   $ \cal M,  {\cal M}$  or   $ \cal M,  \overline {\cal M}$  combinations. 
 The classical interaction between these events  and their BPS  ($\lambda  \rightarrow 0$) limit are:  
    \begin{equation}\label{MonopoleInstanton_Interaction}
    	V_{\mathrm{int}}(r) =  \left\{  \begin{array}{ll} 
	\frac{Q_{M_i} Q_{M_j}}{4\pi r}  (1 - e^{- m_{\phi} r})   \rightarrow   0 & \qquad {\rm for}  \;   (\cal M,   {\cal M})   \cr
		\frac{Q_{M_i} Q_{\overline  M_j}}{4\pi r}  (1 + e^{- m_{\phi} r})    \rightarrow      \frac{2 \pi}{g_3^2} \frac{\a_i \cdot (- \a_j) }{r},    & \qquad {\rm for}  \;   (\cal M,   \overline  {\cal M}) 
	\end{array} \right.
    \end{equation}
    where $r$ is the distance between the two monopole-instantons, $1/r$ is induced by dual photon exchange and  $e^{- m_{\phi} r}/r$ is due to scalar exchange.  Close to BPS limit, where $\lambda  \rightarrow 0$, the interaction between    $ {\cal M},  {\cal M}$  is zero, while the interaction between  
      $ \cal M,  \overline  {\cal M}$ doubles compared to  just photon induced interactions, i.e,  $(1 + e^{- m_{\phi} r}) \approx 2$ for any finite separation $r$. From hereon, we assume   $ \cal M,  {\cal M}$  pairs are non-interacting, while     $ \cal M,  \overline  {\cal M}$  is interacting.   This interaction can be repulsive, attractive or zero depending on charges because 
      \begin{align}
      \alpha_i \cdot  \alpha_j = 2 \delta_{ij} - \delta_{i, j+1} -  \delta_{i, j-1}, \qquad i, j= 1, \ldots, N-1
      \end{align}
        If the interaction is zero, the corresponding duo  is a critical point at any separation (i.e. collection if critical point becomes a critical line), and there are no correlated 2-events.     
      If the interaction is non-zero, this leads to a critical point at infinity and correlated 2-event amplitudes.

    Due to the  non-zero interaction between monopole-instantons, the partition function at 2 instanton order is expressed as 
    \begin{align}\label{PartitionFunction_TwoInstanton}
    	Z_2  &=  \frac{[\C_1]^2}{2!}\int \mrmd^3r_1  \mrmd^3r_2 \, e^{- V_{\mathrm{int}}(| \bf r_1 -  r_2|)} , 
    \end{align} 
    where  
    \begin{align}
     [\C_1] = \x = Ke^{-S_0}  = A \; v^3  \;  S_0^2 \;  e^{-S_0}
     \label{ins-amp}
   \end{align}  
    is the monopole fugacity. $S_0^2$ arises from four zero-modes of monopoles, three  positions, and 1 angular, each contributing   $S_0^{1/2}$, and 
    $v^3 =     v^{4-1}$  arises from the measure   and Jacobian associated with the angular moduli, respectively,  see e.g. \cite{Dorey:1997ij,Fraser:1997xi}.  We will set $v=1$ unless it is useful to see its existence explicitly. Therefore, all lengths are measured in units of monopole radius $r_m= v^{-1}$.

    Defining $\bf R= \frac{1}{2} (r_1 + r_2) $, $ \bf r= r_1 -r_2$, we can express \eqref{PartitionFunction_TwoInstanton} as an integral over the center 
    coordinate and an integral over the relative coordinate. Let us denote  $ \int \mrmd^3R = \V$ where $\V$ is the volume of the space-time manifold.  Then, $Z_2$ takes   the form 
    \begin{align}\label{PartitionFunction_TwoInstanton-2}
    	Z_2  &=   \frac{1}{2}[\C_1]^2 \, \V  \int \mrmd^3r\, e^{- V_{\mathrm{int}}(r)}  
    \end{align} 
    The  integral, $ \int \mrmd^3r\, e^{- V_{\mathrm{int}}(r)}$ as we  will show by using various different regularization, has two parts.  An infinite part  (which can be regularized to   the volume $\V$ of space-time manifold) and a universal finite part, a function of $g^2$. 
    We will show that \eqref{PartitionFunction_TwoInstanton-2}  can be written as 
    \begin{align}\label{subext}
    	Z_2  & \sim   \xi^2  \left( \V^2+     \V      I(g^2) \right)  
    \end{align} 
    This form can be interpreted as a term in the cluster expansion \eqref{EFT2}. 
    In our physical construction, in QFT, the relative position  $r \in \Gamma_{\rm QZM}$   is a  quasi-zero mode coordinate between two instanton event.  However, in semi-classics,  the cluster expansion needs to be suitably generalized to take into account steepest descent cycles over which one needs to integrate over. 
\paragraph*{Repulsive interaction:}
    Let us assume first that the interaction is repulsive and we can write  
    \begin{align}
    V_{\mathrm{int}} = \frac{2\pi}{g^2}\frac{|\a_i\cdot\a_j|}{\tilde r}, \qquad g^2 \equiv \frac{g_3^2}{v}, \; { \rm and } \;\;  \tilde r = v r.
    \end{align} 
    The    integrand  $r^2 e^{-V_{\mathrm{int} }}$ 
    approaches $0$ as $r\rightarrow 0$. For $r \gg (g_3^2)^{-1}$, since  $e^{-V_{\mathrm{int}}} \sim 1$, the integral is divergent and takes the form   $ \int \mrmd^3r \sim \V$. 
    
    To handle the divergence, we first use a hard cut-off in $r$ integration, given by   $R \gg   r_m =v^{-1} $. 	Then, changing the  integration variable as
   \begin{align}
    	z =  \frac{2 \pi |\a_i\cdot\a_j|}{g^2\, \tilde r}  
     \end{align}
 the radial integral is written as
    \begin{equation}    	
    J_2(g^2) = v^{-3} \left(\frac{2 \pi |\a_i\cdot\a_j|}{g^2}\right)^{3} 4 \pi \int_\d^\infty \mrmd z \, z^{-4} e^{-z}.
    \end{equation} 
    where $\delta  =  \frac{2 \pi |\a_i\cdot\a_j|}{  \tilde R g^2 } $. 
    The resulting integral is  
    a representation of the incomplete gamma function:
        \begin{equation}\label{radialIntegral_cutoffResult}
    	J_2(g^2) = 4 \pi v^{-3} \lim_{\delta\rightarrow 0} \left\{ \left(\frac{2 \pi |\a_i\cdot\a_j|}{g^2}\right)^{3} \G(-3,\delta)\right\}.
    \end{equation}	
    Note that $\G(-3,\delta)$ has a branch cut along $\delta =(-\infty,0)$ \cite[Sec. 8.2]{NIST:DLMF}. Its expansion around $\delta=0$, which corresponds to large but finite separation of the monopole instantons, is 
    \begin{equation}\label{IncompleteGamma_Expanison}
    	\G(-3,\delta) \simeq \frac{1}{3\delta^3} - \frac{1}{2\delta^2}+ \frac{1}{2\delta}  + O(\delta)  + 
    	\frac{\ln(\delta) + \g}{6} -\frac{11}{36} 
    \end{equation}
    Using   $\delta  = \frac{2 \pi |\a_i\cdot\a_j|}{  \tilde R g^2} $, we can write $J_2(g^2)$ as
    \begin{align}
    	\label{radialIntegral_cutoffResult2}
    	J_2(g^2) & \simeq \lim_{R \rightarrow \infty}   \frac{4 \pi}{3}  R^3  \left( 1+ O(R^{-1}) \right) +  
    	4 \pi v^{-3}\left(\frac{2 \pi |\a_i\cdot\a_j|}{g^2}\right)^{3}  \frac{1}{6}  \left(  \ln \Big(    \frac{2 \pi |\a_i\cdot\a_j| }{g^2 }     \Big) + \g -   \frac{11}{6}   \right)  
    \end{align}
    as a result of which 
        \begin{align}
    	\label{radialIntegral_cutoffResult3}
    	J_2(g^2)  =  \V + I(g^2)
    \end{align}
    and we get the form $\eqref{subext}$ for $Z_2$.

      { 
    	\noindent
    	{\bf Remark:}  In \eqref{radialIntegral_cutoffResult2},  the leading divergent part  $(  \frac{4 \pi}{3}  R^3  ) $ and
    	the finite part  $I(g^2)$ are independent of the regularizations of integration. We used two other regularization 
    	as well to demonstrate this point. In both,  we obtain  a divergent part (identified as volume) and a finite part  denoted as $I(g^2)$ which agrees with 
	\eqref{radialIntegral_cutoffResult2}. These two other regularizations are described in Appendix \ref{OtherRegularizations}. 

\vspace{0.3cm} 
Let us digest the result we have. Equation \eqref{subext}  is one aspect of critical point at infinity. For example, in the set of action $2S_0$  
configurations, let us consider an  ${\cal M}_{\alpha_i} $  and $ \overline {\cal M}_{\a_{i+1}}$ event. The interaction between them is repulsive so, $\frac{dV_{\rm int}}{dr} =0$ at $r= \infty $.  
 The integral over QZM direction has two parts, a part divergent with volume $\V$ and a finite part $I(g^2)$.  There is also an overall factor of $\V$ that arise from zero mode integration, producing \eqref{subext}.  As a result,  we obtain $e^{-2S_0}( \V^2 + \V I(g^2)) $ from critical point at infinity. The maximally extensive part  in volume provides uncorraleted contribution of two instantons, square of the 1-cluster events  $[\C_1]^2$,   
  while the  sub-extensive part provides correlated 2-instanton contributions, from the set of  2-cluster events  $[\C_2]$. 
    (e.g. $[ {\cal M}_{\alpha_i} \overline {\cal M}_{\a_{i+1}}] \sim  e^{-2S_0}I(g^2) $).  
  
  In this sense, the simplest critical point at infinity contains both  uncorrelated two 1-instanton events and correlated 2-instanton events built in. 
 Then, following the guideline in Appendix \ref{Section: ClusterExpansion}, the equation \eqref{subext} gets the following general form:
  \begin{equation}\label{2instanton_clusterform}
  	Z_2 = \V^2 [\C_1]^2 + \V[\C_2].
  \end{equation}

\paragraph*{Attractive Interactions:} Now, we turn our attention to the more interesting case, i.e. when the sign of $V_{\mathrm{int}}$ becomes negative, and  the  interaction potential	becomes attractive.  
The counterpart of the \eqref{PartitionFunction_TwoInstanton} integral is 
\begin{align}
	\widetilde J_2(g^2) = 4 \pi \int dr \; r^2    e^{+\frac{2 \pi}{g^2}\frac{|\a_i\cdot\a_j|}{r}}
	\label{Jtilde}
\end{align} 
and it can be handled in two related  ways: One is analytic continuation and the other is judicious choice of the integration cycle\footnote{In the past, attractive interaction potentials caused too much confusion and hardship even in the simpler case of quasi-zero mode integral accounting for  instanton anti-instanton interactions  in quantum mechanics. Naive  (and incorrect) perspective on such an integral was following.  If  we consider $e^{- \frac{1}{g^2 r}}$, it vanishes as $ r \rightarrow 0$ very fast,  while $e^{+ \frac{1}{g^2 r}}$ blows up as $ r \rightarrow 0$. Both of the integrands tends to 1 as $r \rightarrow \infty$, which is  responsible for the $\V$ factor that emanates from  the integrals $\int dr r^2    e^{\pm \frac{1}{g^2 r}}$.  But clearly, the attractive interaction seem to pose  a puzzle as $r \rightarrow 0$, as the integrand blows up as $r \rightarrow 0$.   However, this perspective is incorrect in properly done semi-classics. In semi-classical formulation, the integration cycle  is the steepest descent cycle.   For the critical point at infinity,  the descent cycles are 
\begin{align} 
&r \in [0, \infty], \qquad {\rm repulsive}, \cr
& r  \in [0, -\infty] \cup C_{\infty}^{\circlearrowleft},  \;\; {\rm or} \;\;  r  \in [0, -\infty] \cup C_{\infty}^{\circlearrowright},\qquad {\rm attractive},
\end{align}
So, for the attractive case, integration over  $[0, -\infty]$ yields not a pathological integral, but  same integral as in the repulsive case, and the integral over $C_{\infty}^{\circlearrowleft}$ and $C_{\infty}^{\circlearrowright}$ picks the pole  at $ r=0$, depending the orientation producing $\pm i$ in 
\eqref{RadialIntegral_TwoFoldAmbiguity}.  }.  
Here we will use the analytical continuation method, while explain the integration cycles in Appendix \ref{Section: Thimbles_CoulombGas}.
\begin{figure}[h]
	\begin{center}
		\includegraphics[scale=1.5]{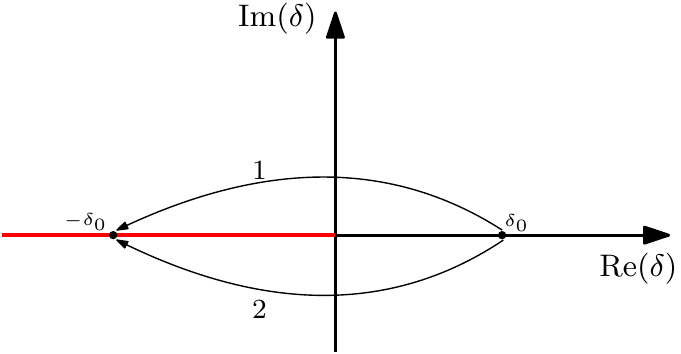}
	\end{center}
	\caption{$\G(-3,\d)$ has a branch cut along the negative real axis. It is possible to compute it at a point $-\delta_0<0$ on the cut by analytically continue its value from $\delta_0>0$. However, paths $1$ and $2$ lead complex conjugate results so that the function $\G(-3,\d)$. This is the source of the ambiguous result of \eqref{RadialIntegral_TwoFoldAmbiguity}. }
	\label{AnalyticalContinuation}
\end{figure}

\vspace{0.3cm}
The analytical continuation idea, which was introduced in the context of QM in  
 \cite{Bogomolny:1980ur},  is  the following. Taking $g^2 \rightarrow -g^2$ in    \eqref{Jtilde} maps the problem to the one studied above, which yields 
 \eqref{IncompleteGamma_Expanison}, $J_2(g^2)$.   Taking $g^2 \rightarrow -g^2$ in  $J_2(g^2)$, therefore, should bring us back to the integral we would like to study,  $ \widetilde J_2(g^2)$. 
However, as we mentioned above, $\G(-3,\delta)$ has a branch cut along the  negative real axis. Since $g^2 \rightarrow -g^2$ corresponds to $\delta \rightarrow -\delta$ in \eqref{radialIntegral_cutoffResult},  there are two possible directions, clockwise  and counter-clockwise for the analytical continuation, see Fig. \ref{AnalyticalContinuation}.  
This freedom of choice leads to two complex conjugate results:  
%
    	\begin{align}\label{RadialIntegral_TwoFoldAmbiguity}
    	\widetilde J_2(g^2) = 	J_2(g^2 e^{\pm i \pi} ) 
    	&= \V - I(g^2)  \pm i  \frac{2 \pi^2}{3}  \left(\frac{2 \pi |\a_i\cdot\a_j|}{g^2}\right)^{3}.
    	\end{align}
    
 Now, as in the repulsive case, let us take an example in the set of instanton and  anti-instanton configurations. If we consider an ${\cal M}_{\alpha_i} $  and $ \overline {\cal M}_{\a_{i}}$ event, the interaction between them is attractive.  The critical point is again at infinity. Now, the integral over the QZM direction has three ingredients,  a part divergent with volume $\V$ and a finite  real part,  and a 
 finite imaginary two-fold ambiguous part.  
 As a result,  the contribution of critical point at infinity to partition function is of the form 
 	    \begin{align}\label{monmon2}
\xi^2 \int \mrmd^3 {\bf r}_1 \mrmd^3 {\bf r}_2 \, e^{- V_{\mathrm{int}}( |{\bf r}_1 - {\bf r}_2|)}  & =   \xi^2  \left( \V^2 -     \V      I_{\pm}(g^2) \right)   
    \end{align} 
and we recover the form of \eqref{subext} and \eqref{2instanton_clusterform}. Note also that the imaginary part belongs to  2-cluster $[\C_2]$  configuration $[{\cal M}_{\alpha_i} \overline {\cal M}_{\a_{i}}]_{\pm} $. 

 \vspace{0.3cm }
   \noindent{\bf Remark:} The overall structure is exactly as in quantum mechanics with instantons. If we compactify $\mathbb R$ to $S^1_\beta$ in a QM system, at second order in the semi-classical expansion, we obtain  $\frac{\xi^2}{2!} \int_{\Gamma} d\tau_1 d\tau_2   \;  e^{- V_{12} } $, where integration over the exact zero mode $ \frac{1}{2}(\tau_1+ \tau_2)$ produce a factor of $\beta$ and the integral over the QZM $\tau_1- \tau_2$ produces $(\beta + I(g^2))$ for repulsive interactions, and $(\beta + I_{\pm}(g^2))$  for attractive interactions,    where $I (g^2)$ is a first order  polynomial in  $\log(1/g^2) $, and  $ I_{\pm}(g^2)=  I(g^2e^{\pm i \pi} ) $.   The final result is  $  \frac{\beta^2 [{\mathcal I}] [{\mathcal I} ] } {2!}  +   \frac{\beta ([{\mathcal I} {\mathcal I}])^1 }{1!} $   or  $  \frac{\beta^2 [{\mathcal I}] [ \overline{\mathcal I} ] } {2!}  +   \frac{\beta ([{\mathcal I}  \overline{\mathcal I}]_{\pm})^1 }{1!} $. The fact that the first term is maximally extensive in $\beta$ and the second term is sub-extensive   is a characteristic feature of critical point at infinity,  and the absence/presence of ambiguity is a feature of the corresponding steepest descent cycle.    As in QFT, in QM, the exponentiation of the cluster expansion yields a contribution to vacuum energy of the form  ${\cal E}_{0, \rm np} \sim   -e^{-S_0}  - e^{-2S_0}I(g^2) - e^{-2S_0}I_{\pm} (g^2) + \ldots$.  

 \subsection{Resurgence structure and predictions for large order behavior}
\paragraph{Vacuum Energy:} As it is defined in \eqref{prolif}, the vacuum energy is related to the terms in the cluster expansion via their exponentiation
$\mcalE  \sim  - \left (  [\C_1] + [\C_2]+  [\C_3] +   [\C_4]  +  \right)$ where   $[\C_k]$ are clusters of $k$-instanton events.  For example, 
\begin{align}
 [\C_2] =  \Big\{  [ {\cal M}_{\alpha_i} \overline {\cal M}_{\a_{i+1}}],    [ \overline {\cal M}_{\alpha_i}  {\cal M}_{\a_{i+1}}],  \; i \in [1, N-2],  \;\;\;   [{\cal M}_{\alpha_i} \overline {\cal M}_{\a_{i}}]_{\pm},  \; i \in [1, N-1]  \Big \}
 \end{align}
We view the Euclidean vacuum as a grand canonical example of all such configurations, not only as a gas of monopoles which takes into account the monopole-instanton events in  $[\C_1]$.   The events in $[\C_2]$ are called magnetic bions and neutral bions\footnote{Although  one may think naively that they are small contributions to leading order monopole effects, in theories with fermions they lead to qualitatively new effects compared to monopoles.  
In QCD(adj) on $\mathbb  R^3 \times S^1$, there are $N$ types of monopoles. Each acquires $2n_f$-fermi zero modes and does not lead to a mass gap for gauge fluctuation. There are $N$-types of magnetic bions and they gap out the dual photons \cite{Unsal:2007jx}.}.
 The non-perturbative contribution to the ground state energy  up to two instanton level comes from  $ [\C_1]$  and   $[\C_2]$  and includes monopole terms of order  $e^{-S_0}$ and bion terms of order  $ e^{-2S_0}$ effects.  Most important for us is the fact that neutral bion contributions are two-fold ambiguous due to the $I_\pm$ factor.  The non-perturbative imaginary ambiguous part is given by:
\begin{align}\label{PartitionFunction_Imaginary}
	\Im   {\cal E}_{\rm np}   &=  \Im [ {\cal M}_{\alpha_i} \overline {\cal M}_{\a_{i}}]_{\pm}  =  \pm   \frac{2 \pi^2}{3}  \left(\frac{2 \pi |\a_i\cdot\a_i|}{g^2}\right)^{3}    [ {\cal M}_{\alpha_i} ]   [ {\cal M}_{ \alpha_{i}}]   \cr 
	&  \sim   \pm   \left(\frac{\mrms_0}{g^2}\right)^{7}  e^{- 2\mrms_0/g^2}.
\end{align}	
where we used vacuum values of the monopole operators \eqref{ins-amp}
 \begin{align}
 \mcalM_{\a_i} = \mcalMbar_{\a_i}  =  \left( \frac{\mrms_0}{g^2} \right)^2 e^{-\frac{\mrms_0}{g^2}}. 
 \label{mon-op3}
 \end{align}
 Note that in \eqref{PartitionFunction_Imaginary}, $7=2+2+3 = 2  \frac{N_{\rm zm}}{2} +N_{\rm qzm}   $ where factors of $2$ are determined by number of zero modes  of instantons divided by two  and    factor of 3 comes from the quasi-zero mode integration, and this will have a crucial effect in large  order growth of perturbation theory. 


   	\vspace{0.3cm} 	
	In  the  Higgsed vacuum \eqref{PerturbativeVacuum}, 
	perturbative contributions to the vacuum energy density $ E_0$ is expected to be an asymptotic divergent series, whose Borel resummation    ${ \cal S}_{\pm} E_0 $ is also two-fold ambiguous. By resurgence, it is expected that the ambiguity in the    ${ \cal S}_{\pm}  E_0 $   to cancel against the ambiguity of the  
$ [ {\cal M}_{\alpha_i} \overline {\cal M}_{\a_{i}}]_{\pm} $, i.e., 
\begin{align}
  \Im [{ \cal S}_{\pm}  E_0  +  [ {\cal M}_{\alpha_i} \overline {\cal M}_{\a_{i}}]_{\pm} ]=0
  \label{res1}
\end{align}
Similarly,   perturbation theory   around the monopole-instanton  
$  \mcalM_{\a_i}  =     ({\mrms_0}/{g^2})^2 e^{-\frac{\mrms_0}{g^2}}  P_{\cal M}  (g^2) $ is also expected to be divergent asymptotic    expansion, whose 
 Borel resummation must have ambiguity at three instanton order.   Indeed, the QZM integration of the 
three instanton event has an ambiguity as well  $ [{\cal M }\overline {\cal M} {\cal M} ]_{\pm} $. 
For semi-classical expansion to be free of pathologies, 
the ambiguous imaginary parts (at three instanton level) must again cancel each other, and one should have 
		\begin{align}
  \Im \left[  [{\cal M }]  { \cal S}_{\pm}  P_{\cal M}  \left( g^2\right)     +  [{\cal M } \overline {\cal M} {\cal M }  ]_{\pm} \right] =0 .
  \label{res-mon-1}
\end{align}

However, it is very difficult to prove   \eqref{res1} and \eqref{res-mon-1}  in full QFT, as we do not have immediate access to large orders of perturbative expansion.\footnote{In general,  the perturbative expansion  is expected to be a double expansion in $g^2$ and $\lambda$.  Indeed, 
 the monopole-action \eqref{action-lambda} depends on both parameters, but the dependence on $\lambda$ becomes very weak in the BPS limit, 
$ \lambda \rightarrow 0$.  We work very close to this limit to avoid such subtleties, and study resurgence properties in terms of coupling  $g^2$  only.}
On the other hand, it is possible to prove  such exact cancellations in quantum mechanics explicitly, e.g, in double-well potential  \cite{Bogomolny:1980ur,  Zinn-Justin:1981qzi,   Pham,  Dunne:2014bca}.

\vspace{0.3cm} 
One may hope that by a dimensional reduction of the Polyakov model, one can prove  the counterparts of \eqref{res1}  \eqref{res-mon-1}  at least in 
quantum mechanical  context. However, it turns out that a naive reduction on $T^2 \times \mathbb R$ with periodic boundary conditions on $T^2$, 
there is a unique perturbative vacuum. Within Born-Oppenheimer approximation, instantons  do not play a role in vacuum properties.   This will be explained in detail in Section.\ref{Section: P_Sector}.  

\vspace{0.3cm} 
On the other hand, by turning on a non-trivial background GNO magnetic flux (for which discrete 't Hooft  flux is non-zero), we can reduce the Polyakov model to QM in such a way that the $N$ perturbative vacuum and $N-1$ types of instantons  of the QFT survive within Born-Oppenheimer approximation.  This  is the merit of turning a  classical  flux background in the Polyakov model.  In the QM  context,   the resurgent cancellations  such as  \eqref{res1} and  
 \eqref{res-mon-1}  are already proven by either using the exact WKB method or by explicit computation.  Below, we make a natural assumption  that such resurgent cancellations survive in QFT as well and make non-trivial predictions for large-order growth of perturbation theory around the perturbative vacuum and monopole-instanton saddle.

\paragraph*{Prediction for large-order perturbative expansion:}
The cancellations in \eqref{res1} and \eqref{res-mon-1} imply that the asymptotic behaviour of the perturbative expansions both around the 
perturbative vacuum saddle   $E_0 = \sum_k   {b}^{(0)}_k g^{2k} $  and  instanton  $ P_{\cal M}  \left( g^2\right)   =  \sum_k   {b}^{(1)}_k g^{2k}$ can be determined by using dispersion relations.  
For example, 
\begin{equation}\label{DispersionRelation_mainText}
b_k^{(0)} = \frac{1}{\pi}\int_0^\infty \mrmd (g^2)\, \frac{\Im  {\cal E}_{\rm np}   (g^2)}{(g^2)^{k+1}} ,  
\end{equation}
where $b_k^{(0)}$ is  perturbative coefficients around the vacuum.  
 Then, using the results for $\Im [\mcalM \overline{\mcalM}]$ in \eqref{PartitionFunction_Imaginary}, we determine the large order behaviour of the perturbative expansions around the vacuum as 
 \begin{equation}
 \label{pred-1}
b_k^{(0)}  
\sim 
 \frac{\G(k+7)}{(2\mrms_0)^k}  \quad ,\quad \mrms_0= 4\pi 
\end{equation}	
 Similarly,  the ambiguity in the three instanton sector is of the form   \eqref{3Instanton_SpectrumAmbiguity}	 
 \begin{align}
\Im [{\cal M } \overline {\cal M} {\cal M }  ]_{\pm}  \sim  \pm \left(\frac{\mrms_0}{g^2} \right)^{12} \ln \left(\frac{\mrms_0  }{g^2}  \right) \, e^{-\frac{3\mrms_0 }{g^2}}  
\label{trip-mon}
\end{align}
Here,   $12= 2+2+2+ 3+3 = 3 \frac{N_{\rm zm}}{2} +2 N_{\rm qzm}    $ where 
 factors of $2$ are determined by number of zero modes  of instantons divided by two
  and factors of threes are 
the consequence of integration over the two quasi-zero mode  (QZM)  coordinates.  The log factor  $   \ln \left(\frac{\mrms_0  }{g^2}  \right)  $  
is also result of the QZM integrations and starts  to appear at third order in semi-classics, and has an interesting  effect on large-order growth.   As one proceeds to  order $p$ in semi-classics, polynomials of order $p-2$ in this logarithmic factor will arise.  Given that 
 instantons 
has prefactors that depend on $g^2$ as $  \mcalM_{\a_i}  =     ({\mrms_0}/{g^2})^2 e^{-\frac{\mrms_0}{g^2}}  P_{\cal M}  (g^2) $, we can determine 
the large order behaviour of perturbation theory around an instanton as:   
\begin{equation}
\label{pred-2}
b_k^{(1)} \sim    \frac{\Gamma(k+10)\,  \ln k }{(2\mrms_0)^k} 
\end{equation}
Again, the factor of ten in Gamma function is just $3  \frac{N_{\rm zm}}{2} +2 N_{\rm qzm}   -  \frac{N_{\rm zm}}{2}$  where the subtraction takes into account the prefactor of instanton.  

Both in \eqref{pred-1} and \eqref{pred-2}, the enhancement relative to $k!$ behaviour is a consequence of the combination of zero and quasi-zero mode contributions. Furthermore, one can check that the extra  $ \ln k$ enhancement in   \eqref{pred-2} is a consequence of  $ \ln \left(\frac{\mrms_0  }{g^2}  \right) $ in \eqref{trip-mon}. Such  $ \ln k$  enhancement of the  large-order behaviour was first shown and tested in the context of quantum mechanics \cite{Dunne:2013ada}. Up to our knowledge, its appearance in a QFT setting is new.    
These predictions can be tested by using stochastic perturbation theory and lattice techniques, see e.g.
 \cite{DiRenzo:2004hhl, Gonzalez-Arroyo:2019zfm}.

\paragraph*{Remark on  mass gap:}
The observation of the cancellation between ambiguous parts is especially important in the light of Polyakov's derivation of the mass gap. Indeed, if we consider the theory on  $\mathbb R^3$, the mass gap would be sourced by the proliferation of the monopoles ${\cal M}_{\alpha_i}$ at leading order, leading to a well-known  result, $m_g^2 \sim e^{-S_0} $. There are also contribution at second order in semi-classics induced by magnetic bions and there are unambiguous $  [ {\cal M}_{\alpha_i} \overline {\cal M}_{\a_{i \pm 1} }] \sim  e^{-2S_0}  $.
However, at third order in semi-classics, there are correction to mass gap sourced by the proliferation of 3-events, e.g.
$  [ {\cal M}_{\alpha_i} \overline {\cal M}_{\a_{i}}  {\cal M}_{\alpha_i}  ]_{\pm}$ which possess two-fold ambiguous imaginary parts. Including both types of effects, 
the result would be $m_g^2 \sim e^{-S_0}  +   e^{-2S_0} + e^{-3S_0} \pm i   e^{-3S_0}   $, which would be quite undesirable.  The implication of \eqref{res-mon-1} is that 
this feature of  semi-classics  would be cured  by a similar feature, sourced by non-Borel summability of the perturbation theory around the instanton. 

The resurgent cancellations  render the combination of semi-classical analysis and perturbation theory  well-defined, even though each part is ambiguous in its own right.  As stated earlier, we cannot currently prove these relations in the context of full QFT on $\mathbb R^3$.  However, we can construct a compactification of the theory on   $ T^2 \times \mathbb R$ in which the instantons of the model of  $\mathbb R^3$ survive by using magnetic GNO (and 't Hooft) flux background. Furthermore, the resurgent cancellation   \eqref{res12}  holds in the quantum mechanical limit. This naturally leads us to conjecture that the relations such as     \eqref{res1} and  \eqref{res-mon-1}  are  valid in the QFT limit on  $\mathbb R^3$.

    	
\section{Turning on  't Hooft flux background}
The original  $SU(N)$ Polyakov model    \eqref{PolyakovModel_Action} has an exact $\mathbb Z_N^{[1]}$ 1-form symmetry, since the dynamical matter field in the theory is in the adjoint representation.  We can consider turning on a classical  background field for the one-form symmetry, called 
 't Hooft flux background  \cite{tHooft:1979rtg, tHooft:1977nqb,vanBaal:1982ag}. 
	On $T^3$, we can insert  $\mathbb Z_N^{[1]}$ fluxes $\ell_{ij}$ through $ij$ faces of the $3$-torus, so we have the freedom to have $N^3$ fluxes. Mathematically, this corresponds to the classification of bundle topologies by    2nd Stiefel-Whitney classes  $w_2 \in H^2(T^3, \mathbb Z_N)$.
	   We will explicitly compactify $\mathbb R^3$ on  $T^2 \times \mathbb R$, and  consider flux $\ell_{12} \equiv \ell$  through $T^2$.

 \vspace{0.3cm}    	
    	Let $\Omega_{\mu}$ denote the transition functions that we use in implementing 't Hooft's twisted boundary conditions connecting adjacent tori. 
    	$\Omega_1(x_2)$ is the transition function between $ (x_1 + L_{1}, x_2) \sim (x_1,  x_2)$, and is 
    	independent of  $x_{1}$, but depends on $x_{2}$  coordinate, and vice versa for $\Omega_2(x_1)$.   We have to impose these boundary conditions  both on gauge field and adjoint scalar, $(A_{\mu}, \f)$. For simplicity, let us write the explicit construction for $\f$. 
    	Impose
    	\begin{align}
    		\f(x_1 +L_1, x_2)&= \Omega_1(x_2)  \f(x_1,  x_2)   \Omega_1^{-1}(x_2), \qquad  \cr
    		\f(x_1 , x_2 + L_2)&= \Omega_2(x_1)  \f(x_1,  x_2)  \Omega_2^{-1}(x_1)
    		\label{flux}
    	\end{align}
    	We can connect  the fields at the corners,  $\f(x_1 +L_1, x_2+L_2)$ with $\f(x_1,  x_2)$,  via two different paths: 
    	$(x_1, x_2) \rightarrow 
    	(x_1 +L_1, x_2)  \rightarrow (x_1 +L_1, x_2+L_2)$ or  replacing middle point with $(x_1, x_2+L_2)  $.  
     Consistency  of the field at  $\f(x_1 +L_1, x_2+L_2)$ demands the transition matrices   $\Omega_{\mu}$ to obey 
    	\begin{align}
    		\Omega_1(L_2) \Omega_2(0)&= \Omega_2(L_1) \Omega_1(0)  e^{i  \frac {2 \pi \ell }{N} } 
    		\label{flux2}
    	\end{align}
    	where  $\ell=0,1, \ldots, N-1$ is the  't Hooft flux through $12$.    Under local gauge transformations, obviously,  $\f(x_1, x_2) \longrightarrow  g(x_1, x_2) \f(x_1, x_2) g(x_1, x_2)^{\dagger}$. Therefore, the transition function must also transform for the gauge covariance of \eqref{flux}.  The gauge invariant  information in transition matrices   \eqref{flux2} is just 
    	$\ell$, the discrete  flux modulo $N$.

    	\vspace{0.3cm} 
    	It is possible to  undo twisted boundary conditions on $T^2 \times \mathbb R$,  and replace them with a background 
    	$B^{(2)}$  fields associated with 1-form center symmetry $\mathbb Z_N^{[1]} $ given by 
    	\begin{align}
    		\frac{N}{2\pi}  \int_{12} B^{(2)}  =  \ell_{12} = \ell  \;\;\;  {\rm mod}  \;\; N
    	\end{align}
	In the abelianized theories, it is also easy to relate the discrete 't Hooft flux and magnetic  GNO fluxes \cite{Unsal:2020yeh}. 
    	The magnetic  GNO   flux  through 
    	the 12-surface can be written as
    	\begin{align}
    		\int_{12}  \bm B = \frac{2 \pi} {g}  \bm \mu_\ell,   \qquad \bm  \mu_\ell \in  \Gamma_w^{\vee}
    	\end{align}
	where $\ell$  is a non-zero  magnetic $N$-ality,  valued in  $  \Z_N  \cong  \Gamma_w^{\vee} / \Gamma_r^{\vee}$.
    	Note that the dynamical magnetic  monopole charges in the theory are associated with the co-root lattice $\alpha \in \Gamma_{r}^{\vee}$ and do not change magnetic $N$-ality.   Therefore, the GNO-flux  configurations  
    	$    \mu_1,  \mu_1 -\alpha_1,  \mu_1 -\alpha_1 -\alpha_2, \ldots $ are all associated with $\ell=1$ discrete  flux.\footnote{Recall that  charges with non-zero $N$-ality  are not present in the $SU(N)$ theory, but we can insert a  GNO flux through $T^2$  to probe the dynamics of the theory.    Even if we gauge $\mathbb Z_N^{[1]} $ completely and move to  {\it pure} $PSU(N)$ theory (without magnetic matter),  charges with non-zero magnetic $N$-ality are still not present dynamically, but can be introduced as probes.  In  more general $PSU(N)$ theories,  they can be introduced as dynamical 
	magnetic matter fields. }

\vspace{0.3cm} 
And finally,  we may think the $\ell $-units  't Hooft flux as a  non-dynamical center-vortex \cite{Greensite:2003bk}. The holonomy of the gauge field around the 't Hooft flux insertion  is a center element,  $e^{i \oint a} = e^{i  \frac{2 \pi \ell}{N}}$.    	
    	
\subsection{Formal construction of coupling to  $\mbbZ_N$ TQFT}\label{CouplingTQFT}
We continue our discussion by coupling the $SU(N)$ Polyakov model to a $\mathbb Z_N$ topological gauge theory.   The resulting theory is     	$SU(N)/\mathbb Z_N = PSU(N)$ gauge theory. Recall that there is  no discrete theta angle in 3d gauge theory,  hence, we only discuss one type of $PSU(N)$ theory. Coupling to $\mathbb Z_N$ TQFT does not alter the local dynamics in the Polyakov model, but the spectrum of line operators and relatedly, the periodicity of dual photon field changes. In particular, 't Hooft operators with magnetic charges belonging to the co-weight lattice are useful operators in $PSU(N)$ and Wilson loops sourced by representation belonging to weight lattice are  no longer genuine line operators. 
	
\vspace{0.3cm} 
To introduce a   background gauge  field  for the  $\Z_N^{[1]}$  1-form symmetry, one introduces a  pair of $U(1)$ 2-form and 1-form  gauge fields $(B^{(2)}, B^{(1)}) $ satisfying \cite{Kapustin:2014gua, Gaiotto:2014kfa}
\begin{align}
 N B^{(2)}= \diff B^{(1)},  \qquad N \int B^{(2)}=  \int \diff B^{(1)} = 2 \pi \Z.
 \label{ZN-YM}
\end{align}
In the standard 't Hooft construction, this amounts to saying that 
\begin{align}
 \frac{N}{2\pi}  \int_{\Sigma ij} B^{(2)}  =  \ell_{ij}  \;\;\;  {\rm mod}  \;\; N ,
\end{align}
where $\ell_{ij}$ is 't Hooft flux through the 2-cycle $\Sigma_{ij}$  of the three torus $T^3$. 
    	
	\vspace{0.3cm} 
    It is convenient to express the action of the $\Z_N$ TQFT as
    \begin{align}
    	Z_{{\rm top}} &= \int  \Diff   B^{(2)} \Diff B^{(1)}    \Diff  C^{(1)} \;  e^{ i \int C^{(1)}  \wedge ( N B^{(2)}-\diff B^{(1)})  }  ,
    		\label{top-3}
    \end{align}
    where $C^{(1)}$ is Lagrange multiplier.  Equation \eqref{top-3} is invariant under the 1-form gauge  transformation, $
    B^{(2)}\mapsto B^{(2)}+\diff \Lambda^{(1)},  \;\;  B^{(1)}\mapsto B^{(1)}+N\Lambda^{(1)}$.      
 
 	\vspace{0.3cm}    
    To couple the 3d gauge theory to the background  $B^{(2)}$ field,  we first promote the  $SU(N)$ gauge field $A$ into a $U(N)$ gauge field  $\tilde A$ given by $\widetilde{A}=A+{1\over N}B^{(1)}$.  As a result, under 1-form gauge  transformation, $ \widetilde{A}\mapsto \widetilde{A}+\Lambda^{(1)},   \;   \widetilde{F} \mapsto  \widetilde{F} + \diff \Lambda^{(1)} $	i.e., the field strength is not  gauge invariant under 1-form gauge transformations.  The invariant combinations  are $ \tG   = \widetilde{F}-B^{(2)}$, and  $\tilde \mcalD = \mrmd - \tA - \frac{1}{N} B^{(1)}$. 
    	Then, Polyakov model  \eqref{PolyakovModel_Action} in the   $(B^{(2)}, B^{(1)}) $ background  field becomes: 
    	\begin{equation}\label{CoupledAction}
    		\mathcal{S}[ (B^{(2)}, B^{(1)}) ]= \frac{1}{2 g^2}\int \left( \tr \, \tG\wedge \*\tG + \tr \, \tilde\mcalD \f \wedge \*\tilde \mcalD\f + V(\f) \right) ,
    	\end{equation}
    	which is invariant under both 1-form and 0-form gauge transformations. 
    	To obtain the 
    	the partition function of the  $SU(N)/\Z_N$ gauge theory, we also need to integrate over   the   2-form gauge fields:
    	\begin{align}
    		Z_{PSU(N)} =  \int  \Diff   B^{(2)} \Diff B^{(1)}    \Diff  C^{(1)}  \Diff  \widetilde A \;  
    		e^{ i \int C^{(1)}  \wedge ( N B^{(2)}-\diff B^{(1)}) }   e^{-S[ B^{(2)}, B^{(1)},  \widetilde A] }
    		\label{top-5}
    	\end{align}
    	This concretely  relates  the partition functions of $PSU(N)$ theory and $SU(N)$  theories as:
    	\begin{align}
    		Z_{PSU(N)} = \sum_{\ell_{ij} \in  \mathbb Z_N^3} Z_{SU(N)}(\ell_{ij}) 
    	\end{align}
    	where $Z_{SU(N)}(\ell_{ij}) $ is the partition function of the $SU(N)$ theory in the  $(\ell_{12}, \ell_{23}, \ell_{31}) 
    	\in \mathbb Z_N^3$ discrete flux background.

    	\section{QM reduction with instantons in Born-Oppenheimer limit}
	\label{Section: CoupledTheory_IR}
    	In this section, we focus on the  properties of the quantum mechanical system on small $T^2 \times \mathbb R$ without and with 't Hooft flux. 
    	In the large volume  limit where $T^3$ is much larger than the correlation length $m_g^{-1}$, there is no distinction in the local dynamics examined via    $Z_{SU(N)},  Z_{SU(N)}(\ell_{ij})$  and  $Z_{PSU(N)}$.  
	 However, in small $T^2 \times \mathbb R$, a quantum mechanical limit,  there is much more benefit in studying the dynamics of the quantum theory via  $Z_{SU(N)}(\ell_{ij})$ then $Z_{SU(N)}$. In particular, in the Born-Oppenheimer approximation,  $Z_{SU(N)}$ does not remember the instantons of the theory on $\mathbb R^3$ in the small   $T^2 \times \mathbb R$ limit, 
    	while  the instanton configurations  and degenerate harmonic vacua survive in $Z_{SU(N)}(\ell_{12} \neq 0 )$.  We explain this structure below. 
    	%
    	
    	\subsection{Discrete 't Hooft and magnetic GNO  flux, and Flux states }\label{Section: P_Sector} 
    	There are two fluxes that one can turn on for gauge theory formulated on $T^2 \times \mathbb R$.  These can be identified as:
	\begin{itemize} 
\item Magnetic GNO flux,  $ \bm \mu_\ell \in  \Gamma^{\vee}_{w}$
    \item Discrete 't Hooft flux, $\ell \in  \Z_N  \cong   \Gamma_w^{\vee} / \Gamma_r^{\vee} $
    \end{itemize}
  These two fluxes are not independent. 
 The   't Hooft flux $\ell_{12} $ can be viewed as 
    	being classified by $\Z_N$ that describes   the quotient   $\Gamma^{\vee}_{w}/  \Gamma^{\vee}_{r}$.  On the other hand, the GNO flux takes possible values in $\Gamma^{\vee}_{w}$. 
    
        	\begin{figure}[t]
    		\begin{center}
    			\includegraphics[scale=0.3]{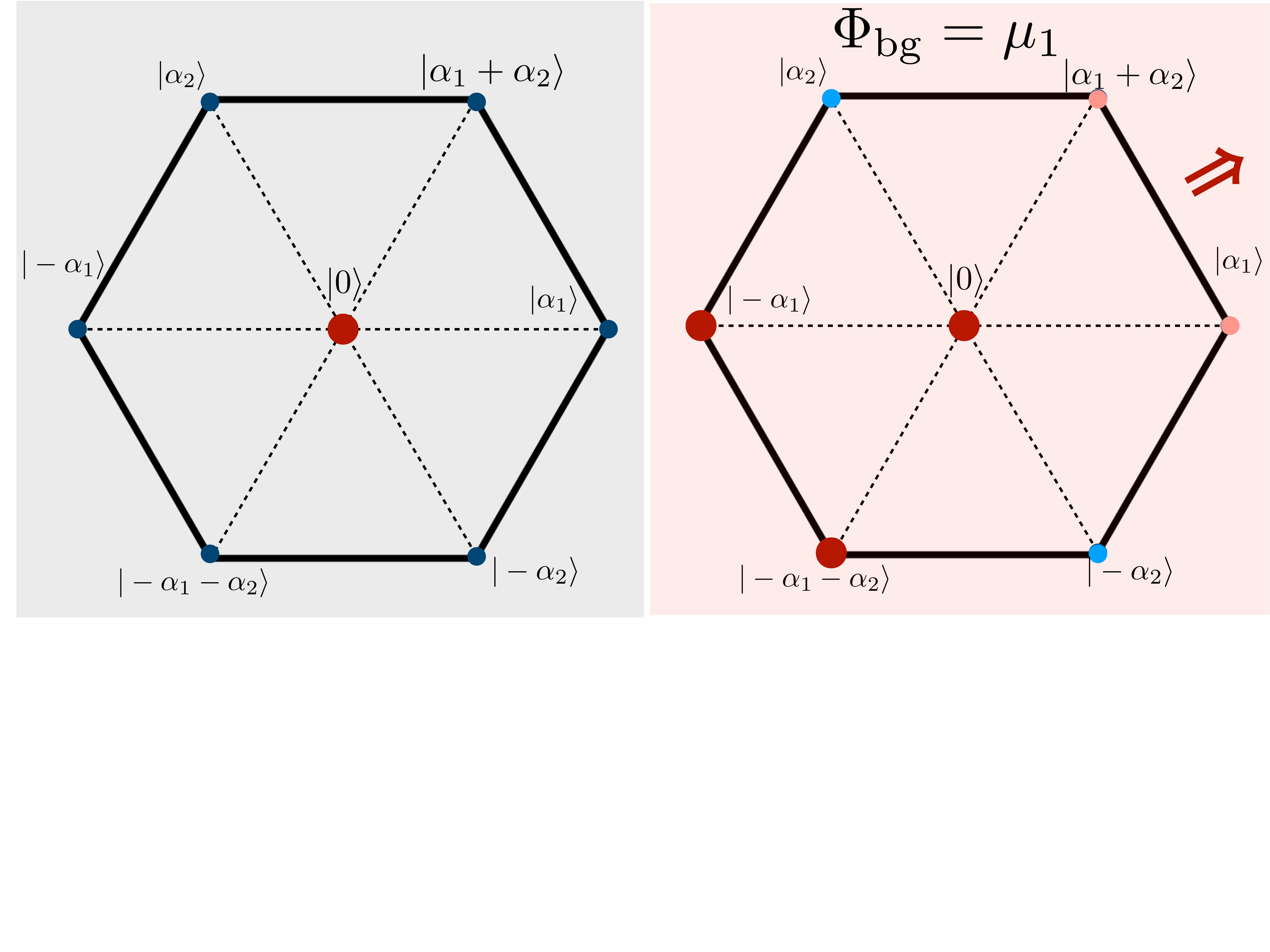}
    		\end{center}
    		\vspace{-3.3cm}
    		\caption{Compactification of $ SU(3)$ Polyakov model to QM  for $\Phi_{\rm bg} =0$   and  $\Phi_{\rm bg} = \mu_1
    			$   background magnetic flux. For $\Phi_{\rm bg} =0$,  the vacuum is unique perturbatively  and flux states have higher energy.  For  $\Phi_{\rm bg} = \mu_1$, the vacua is three-fold degenerate perturbatively, and instanton persist 
    			in the quantum mechanical small  $A(T^2)$ limit.  }
    		\label{wowy2}
    	\end{figure}

    	Considering the system on $T^2 \times \mathbb R$, let us express the magnetic GNO  flux passing through $T^2$   as 
    	\begin{equation}\label{Flux_Monopole}
    		{\bm \F}  = \int_{12}  \bm B =  \bm B   A_{T^2} =   \frac{2 \pi} {g}  \bm \mu_\ell ,
    	\end{equation}
    	where $A_{T^2}$ is the area of $T^2$. 
    	The energy of the flux configuration is given by 
    	\begin{equation}\label{Energy_MonopoleState}
    		E  = \frac{1}{2}\int_{T^2} \bm B^2 = \frac{1}{2 A_{T^2}}   {\bm \F}^2 .
    	\end{equation}
    	The configuration with minimal energy  does not have any flux, $\Phi=0$. In the Polyakov model , the dynamical instantons changes the flux by 
    	$\Delta  {\bm \F} = \frac{2 \pi}{g} \alpha$, and the energy of such configuration is  $E=  \frac{1}{2 A_{T^2}}    \frac{8 \pi^2}{g^2} $.  This is  much higher than perturbative vacuum with $E_0 =0$. Therefore, in the QM limit  of the  Polyakov model, perturbative vacuum is unique.  Only in the limit 
    	$ A_{T^2} \rightarrow \infty $, different flux sectors become degenerate \cite{Banks:2008tpa, Tong}. See Fig.~\ref{wowy2}, left for $SU(3)$.  Perturbative vacuum is unique  and first excited state is six fold degenerate.

   Now, let us check what happens once we turn on a classical background flux. 
  In that case, we can write the flux of a configuration as  
    	$(\F_{\rm bg} + \F)$, a background plus a dynamical part. This makes things more interesting. The energy of such configurations is: 
    	\begin{equation}\label{Energy_MonopoleState_bg}
    		E =  \frac{1}{2 A_{T^2}}   { ( \bm \F_{\rm bg} + \F)   }^2
    	\end{equation}
    	The point is, even when $\F=0$, energy associated with $ \bm \F_{\rm bg} $ is positive. As such, changing  flux by $\alpha \in \Gamma_r^\vee $ has a standing  chance to be degenerate with $ \bm \F_{\rm bg} $ configuration. And indeed, this is the case.     
    	
    	Assume  $\bm \F_{\rm bg}  =   \frac{2 \pi} {g}  \bm \mu_1$. Then, it is easy to see that the following $N$  magnetic flux configurations 
    	\begin{align}
    		{\bm \nu}_1\equiv {\bm  \mu}_1,  \;\;  {\bm \nu}_2 \equiv {\bm  \mu}_1 -   {\bm  \alpha}_1,  \;\;  {\bm \nu}_3 \equiv {\bm  \mu}_1 -   {\bm  \alpha}_1 -    {\bm  \alpha}_2,    \;\; \ldots   \;\;  
    		{\bm \nu}_N \equiv {\bm  \mu}_1 -    \sum_{a=1}^{N-1} {\bm  \alpha}_a, 
    		\label{magneticflux-2} 
    	\end{align}
    	are  degenerate classically.  The   energy of these  flux states are given by  
    	\begin{equation}\label{Energy_MonopoleState}
    		E_i = \frac{1}{2}\int_{T^2} \bm B^2 = \frac{1}{2 A_{T^2}} \left(\frac{2\pi  \bm \nu_i}{g}\right)^2 =\frac{2 \pi^2}{g^2 A_{T^2}} \left(1 - \frac{1}{N} \right),
    	\end{equation}
    	where we used the fact that weights of defining representation  $\bm \nu_i$ have constant length $ \bm \nu^2_i = \left(1 - \frac{1}{N} \right)$.  Note that the other flux states which differ from the set  \eqref{magneticflux-2}  by  roots necessarily have higher energies by factors of $\frac{1}{g^2 A_{T^2}}$ and are ignored within Born-Oppenheimer approximation.  As shown in  Fig.~\ref{wowy2}, right for $SU(3)$,  perturbative vacua  is now three-fold degenerate. 
    	
    	It is also important to note that since magnetic  $N$-ality is defined modulo roots, these tunneling can never change 
    	the   discrete flux of the background, which is given by  $\ell=1$.  To construct higher  discrete  flux  states, one needs to start with a background flux  $\bm \mu_{\ell} $ corresponding to a higher   weight representation.  
    	
    	Let us denote the harmonic states in quantum mechanics by their   flux configurations   
    	\begin{align}
    		\Big| {\bm \nu}_j \Big \rangle   = \Big|  {\bm  \mu}_1 -    \sum_{a=1}^{j-1} {\bm  \alpha}_a \Big \rangle 
    	\end{align}
    	The simplest tunnelings are between the  nearest neighbor states  $| {\bm \nu}_j \>  \rightarrow | {\bm \nu}_{j + 1} \>  $.  We identify the changes in  magnetic flux  and  discrete  flux  as: 
    	\begin{align}
    		\Delta &    \int_{T^2}  {\bm B} =     \frac{2 \pi} {g}   \left( {\bm \nu}_j  - {\bm \nu}_{j+1} \right) = -   \frac{2 \pi} {g} {\bm \alpha}_a, 
    		\qquad a =1, \ldots, N-1.  \cr
    		\Delta  & \left(  \frac{N}{2\pi}  \int_{12}  B^{(2)}  \right) =  0   \;\;  
    	\end{align}


    	\subsection{Tunneling between flux states and leading instanton effects} 
	\label{sec:tunnel}
In $\ell=1$ unit of discrete flux backgrounds,  $N$-states given in \eqref{magneticflux-2} are degenerate and separated from higher states by a large perturbative   gap.       	In Born-Oppenheimer approximation,  we can focus on  these lowest $N$ states split from each other only by tiny non-perturbative factors. 
The degeneracies between these    $N$ states can  lift via the tunneling effects.  We can consider the Hamiltonian in tight-binding approximation, where only hopping between nearest-neighbor configurations are taken into account. 
    	Let us write 
    	\begin{align}
    		H= E_{\rm bg} { \bf 1}_N + \widetilde  H ,
    	\end{align}
    	where $\widetilde  H$ accounts for the leading instanton effects.  Let us write first  few cases explicitly: 
    
    \vspace{0.3cm}
    \noindent
    	{\bf N=2}:  The QM reduction of the  $SU(2)$  Polyakov model in the BO approximation is same as  the double-well potential shown in Fig.\ref{wowy}, right.  To describe the low end of the spectrum, we can use  the tunneling Hamiltonian: 
    	\begin{align}\label{tunnel2}
    		\widetilde  H =   -
    		\left[
    		\begin{array}{cc}
    			0 & \xi  \\
    			\xi & 0  \\
    		\end{array}
    		\right]
		\end{align}
    	where $ \xi  =K e^{-S_0} $ is the instanton fugacity. 
    	The ground state  and first excited state are symmetric/anti-symmetric combinations of  fractional  flux   states   $|\pm \frac{1}{2} \rangle   $:  
    	\begin{align}
    	|\Psi_0 \rangle & =  
    		\textstyle {  \frac{1}{\sqrt 2} }\left( | \textstyle  \frac{1}{2}  \rangle  +  |  \textstyle { - \frac{1}{2} } \rangle \right), \cr
		 	|\Psi_1 \rangle & =  
    		\textstyle {  \frac{1}{\sqrt 2} }\left( | \textstyle  \frac{1}{2}  \rangle  -  |  \textstyle { - \frac{1}{2} } \rangle \right), 
    	\end{align}
    	with energy eigenvalues $\tE_0= -\xi$ and  $\tE_1= +\xi$. The gap  is given by an instanton factor, $\Delta E = 2 \xi$ at leading order in semi-classics.

       \vspace{0.3cm}
    \noindent 	
    	{\bf N=3}:  The QM reduction of the  $SU(3)$ theory in the BO approximation is  the triple-well potential  on  $\mathbb R^2$, shown in Fig.\ref{wowy2}, right.  Low energy spectrum can be described by the  tunneling Hamiltonian: 
    	\begin{align}\label{tunnel3}
    		\widetilde  H =   
    		- \left[
    		\begin{array}{ccc}
    			0 & \xi  & 0 \\
    			\xi  & 0 & \xi  \\
    			0 & \xi  & 0 \\
    		\end{array}
    		\right]
    	\end{align}  
	Note that the transition amplitudes $ \langle {\bm \nu}_2| e^{-\beta H}  |{\bm \nu}_1  \rangle=   \langle {\bm \nu}_3| e^{-\beta H}  |{\bm \nu}_2  \rangle =\xi $ is an instanton effect, but  $ \langle {\bm \nu}_3| e^{-\beta H}  |{\bm \nu}_1  \rangle \sim \xi^2$ is higher order in semi-classics, hence not included at leading order 	analysis.\footnote{This is one of the main distinctions  between   deformed Yang-Mills  \cite{Unsal:2008ch} and Polyakov model \cite{Polyakov:1976fu}, (or  group vs. algebra valued adjoint scalar fields). In the former,  
	 $ \langle {\bm \nu}_3| e^{-\beta H}  |{\bm \nu}_1  \rangle = \xi$ as well, and the affine monopole with magnetic charge $\alpha_N$ is on the same
	 footing with the rest of monopoles  associated with the simple root system  $\alpha_1, \ldots,  \alpha_{N-1}$. }  
    	The energy eigenvalues  are $ \tE_0= -\sqrt 2 \xi,  \; \tE_1= 0, \;  \tE_2= \sqrt 2 \xi$ at leading order in semiclassical expansion.  The corresponding eigenstates are 
    	\begin{align}
    		|\Psi_0 \rangle &= \textstyle{ \frac{1}{2}  \left( |{\bm \nu}_1 \rangle + \sqrt 2  |{\bm \nu}_2 \rangle + |{\bm \nu}_3 \rangle  \right),} \cr
    		|\Psi_1 \rangle  & =  \textstyle{ \frac{1}{ \sqrt 2}  \left(  |{\bm \nu}_1 \rangle  -  |{\bm \nu}_3 \rangle  \right), }  \cr
    		|\Psi_2 \rangle &=  \textstyle{ \frac{1}{2}  \left( |{\bm \nu}_1 \rangle - \sqrt 2  |{\bm \nu}_2 \rangle + |{\bm \nu}_3 \rangle  \right)}
    	\end{align}
    	%
    	The degeneracy is lifted at 1-instanton order and the energy gap is  an instanton factor.

    	       \vspace{0.3cm}
    \noindent 	
    	{\bf N=4:} The $ SU(4) $   gauge theory reduces to a quantum mechanical system on $ \mathbb R^3$ with four degenerate minima.  The tunneling Hamiltonian takes the form:
	\begin{align}
  	\widetilde  H =    -   	\left[
    	\begin{array}{cccc}
    	0 &  \xi & 0 & 0 \\
    	\xi & 0 &  \xi & 0 \\
    	0 &  \xi  & 0 &  \xi \\
    	0 & 0 &  \xi & 0 \\
    	\end{array}
    	\right]
    	\end{align}
    	with eigenfunctions
	 	\begin{align}
    	|\Psi_0 \> &=\textstyle{ \mathcal{N} \left( + |{\bm\n }_1\> + \frac{1+\sqrt{5}}{2}  | {\bm \nu}_2 \> + \frac{1 + \sqrt{5}}{2}|{\bm\n }_3\>+|{\bm\n }_4\>\right) }\nonumber \\ 
    	|\Psi_1 \> &= \textstyle{ \mathcal{N} \left(- |{\bm\n }_1\> + \frac{1-\sqrt{5}}{2} | {\bm \nu}_2 \> -\frac{1-\sqrt{5}}{2}|{\bm\n }_3\>+|{\bm\n }_4\>\right) }\nonumber \\ 
    	|\Psi_2 \> &= \textstyle{  \mathcal{N} \left( + |{\bm\n }_1\> + \frac{1-\sqrt{5}}{2}| {\bm \nu}_2 \> + \frac{1-\sqrt{5}}{2}|{\bm\n }_3\>+|{\bm\n }_4\>\right) }\nonumber \\ 
    	|\Psi_3 \> &= \textstyle{ \mathcal{N} \left(-|{\bm\n }_1\> + \frac{1+\sqrt{5}}{2}| {\bm \nu}_2 \> -\frac{1+\sqrt{5}}{2}|{\bm\n }_3\>+|{\bm\n }_4\>\right) }
    	\end{align}
    	where $\mathcal{N}  $ 
	is the normalization constant. The energy eigenvalues are $\tE_0 = -\frac{1+\sqrt{5}}{2}\xi $, $\tE_1 = \frac{1-\sqrt{5}}{2}\xi$, $ \tE_2=\frac{-1+\sqrt{5}}{2}\xi$ and $\tE_3= \frac{1+\sqrt{5}}{2}\xi$.  The gap is again at 1-instanton order.\\
	
	Note that in all cases, the eigenstates are a linear combination of the fractional  flux states. Because of the background GNO flux insertion, all flux states are in 
	co-weight lattice, but the important point is that the separation between the flux states is in the co-root lattice, corresponding to dynamical monopole-instantons in the theory.

\subsection{Resurgence in QM reduction of Polyakov model with 't Hooft flux}
The $SU(2)$  gauge theory with $\ell_{12}=1$ unit of 't Hooft flux  reduces to double-well quantum mechanics, as shown in Fig.~\ref{wowy}. The working of resurgence in the double-well potential is well understood by now.  Since this knowledge is present in literature \cite{Zinn-Justin:1981qzi, Dunne:2014bca} we state  here the main implication for our analysis. 
    	
\begin{itemize}
\item
Perturbation theory around the degenerate harmonic flux vacua  $|\pm \frac{1}{2} \rangle $ is an divergent asymptotic expansion, \[E_0 = \sum_k{\cal E}^{(0)}_k g^{2k}.\] This series  is non-Borel summable.  The Borel resummation ${ \cal S}_{\pm}  E_0$    is   two-fold ambiguous. Similarly, the instanton-antiinstanton configuration is also  two-fold ambiguous.  The two   ambiguities  cancel exactly. 
		\begin{align}
  \Im \left[{ \cal S}_{\pm}  E_0  +  [{\cal I } {\cal I} ]_{\pm} \right] =0
  \label{res11}
\end{align} 
where $\cal I$ is the magnetic flux changing instanton event  
(
that extrapolates from $ \Phi=- \frac{1}{2}$ to $ \Phi=+ \frac{1}{2}$.  This  is the dimensional  reduction of the  monopole ${\cal M}$ 
 in the original Polyakov model obtained  by using method of replica.  The mapping is formulated in Appendix~\ref{sec:mapping-MI}. 
 By the appropriate scale separations in the  geometric construction, the action of the  instanton  in QM  must be the  same as the action of the monopole in QFT,  $S_{ \cal I} = S_{ \cal M}$.  In fact, in dYM,  on $\mathbb R \times T^2 \times S^1$,  we can prove this statement rigorously,  that $S_{ \cal I} = S_{ \cal M} = \frac{1}{N}  \frac{8 \pi^2}{g^2} $ due to the fact that  we can turn on two types of discrete fluxes which guarantees that  topological charge is $Q= 1/N$. Since  monopole  is a solution to BPS equation with the appropriate boundary condition, this show that these configuration in QM limit of dYM have  fractional action.  

\item Exact WKB and exact quantization methods prove implicitly that this type of cancellations occurs to all  non-perturbative orders, repeating itself around all non-perturbative saddles.  See  \cite{DDP, Sueishi:2020rug}. 
 In particular,  all  orders  perturbative fluctuations around the instanton $E^{(1)} \sim e^{-S_0} P_I (g^2) $ can be determined by using P/NP relation  \cite{Alvarez, Dunne:2013ada}
 and   $P_I (g^2)$ is also a divergent asymptotic expansion.    
   The exact  quantization conditions implicitly prove that the ambiguous imaginary parts (at three instanton level) again cancel each other:
		\begin{align}
  \Im \left[  [{\cal I }]  { \cal S}_{\pm}  P_I \left( g^2\right)     +  [{\cal I } \overline{\cal I} \I ]_{\pm} \right] =0 .
  \label{res12}
\end{align}     	
	\end{itemize}	
The cancellations in \eqref{res11} and \eqref{res12} are  proven   in quantum  mechanics.   Since the instantons in quantum mechanics are mappings of  monopoles in QFT compactified on $T^2 \times \mathbb R $ with the insertion of  't Hooft flux  
(which guarantees that instanton survives in QM limit), this leads  us to conjecture that the counterpart of these resurgent relations,    \eqref{res1} and \eqref{res-mon-1},   hold in QFT. This leads to some predictions,  \eqref{pred-1} and \eqref{pred-2}  in QFT, that can be  tested by lattice simulations.

 \section{Polyakov vs. deformed YM  with  't Hooft flux} 
 \label{compare}
 Polyakov model  on    $\mathbb R^3$  and deformed YM  (dYM) on small   $\mathbb R^3 \times S^1$ are intimately related, but different theories. Here, we would like to provide a succinct comparison of the properties of these two theories. But first, let us briefly remind basic aspects of dYM for completeness.  
 
    \vspace{0.3cm}
  Deformed YM theory is the center-stabilizing double-trace deformation of YM theory on small  $\mathbb R^3 \times S^1$ \cite{Unsal:2008ch}.  The gauge holonomy   $U $ around $S^1$ plays the same role as the Higgs field in Polyakov model. However, in Polyakov model, $\f$ is algebra valued, while in dYM, 
  $U $ is group valued.  In the vacuum,   $U$ acquires a vev $U = {\rm Diag}  \left( e^{i v_1},  e^{i v_2}, \ldots,  e^{i v_N} \right), \;  |v_{i+1} - v_i|= 
  \frac{2 \pi}{N} $ and the theory  undergoes dynamical abelianization $SU(N) \rightarrow U(1)^{N-1} $. 
  Due to the compactness of the gauge holonomy  in dYM, 
 there exists an affine monopole associated with the affine root $\alpha_N$ which has the same action as the other monopoles 
 \cite{Lee:1997vp, Kraan:1998sn}.  
Even though there are fundamental differences between these two  theories that start with their symmetries or even parameters that one can write in their action,  as well as their global structures,  their local dynamics are extremely similar.  

\subsection{Symmetries and 't Hooft fluxes}   
 First, let us start with  formal differences.   For Polyakov model, the center symmetry is  $ (\Z_N^{[1]})_{\rm E} $ while  for dYM  on $\mathbb R^3 \times S^1$, the center-symmetry is $ (\Z_N^{[1]})_{\rm E}  \times  (\Z_N^{[0]})_{\rm E} $  which descends from $ (\Z_N^{[1]})_{\rm E}   $ on $\mathbb R^4$: 
   \begin{align} 
 SU(N) \;  \;\; {\rm P:} \qquad   &  (\Z_N^{[1]})_{\rm E}   \cr 
SU(N)  \;   \;\;   {\rm  dYM:} \qquad  &   (\Z_N^{[1]})_{\rm E}  \times  (\Z_N^{[0]})_{\rm E}
\label{specth}
  \end{align}  
  The charged operators under 1-form symmetry are Wilson line operators and under the 0-form symmetry, it is Polyakov loops.  

     \vspace{0.3cm}
 In Polyakov model defined on $T^2 \times S^1_\beta$,  we can  turn on $N^3$  discrete  fluxes  valued in  $H^2(T^3, \mathbb Z_N)$ and discrete theta angle is absent.   In dYM on $T^3 \times S^1_\beta$,  one can turn on $N^6$ discrete  fluxes  valued in  $H^2(T^4, \mathbb Z_N)$ and there is also a discrete theta angle. Finally, gauging  the  center symmetry,  i.e., summing over all possible fluxes gives us $PSU(N) $  theories.   The partition function for the $PSU(N)$ theories can be written as  
 	\begin{align}
  {\rm P:} \qquad    	&	Z_{PSU(N)} = \sum_{\ell  \in  \mathbb Z_N^3} Z_{SU(N)}(\ell )  \cr
{\rm  dYM:} \qquad  &Z_{PSU(N)_p} = 
  \sum_{ \substack{W \in \Z \\  \ell, m  \in \Z_N^3} }     e^{  i  \frac { 2 \pi }{N} p  \; (\ell \cdot m)}      e^{ i \theta \left( W+\frac{ (\ell \cdot m)}{N}  \right) } Z_{W}( \ell, m)      \qquad  \qquad 
    	\end{align} 
There is no theta angle in the Polyakov model, but since deformed YM is a  locally 4d theory,  there is a topological theta angle,  and an  associated topological charge quantized in integer units in $SU(N)$ theory.  In $PSU(N)$, topological charge is quantized in units of $1/N$. The implication of this for the $SU(N)$ dynamics  is discussed in \cite{Unsal:2020yeh}. 
	
\vspace{0.3cm} 
		 Gauging  
		  the $ (\Z_N^{[1]})_{\rm E}  $  1-form symmetry in the confining  $SU(N)$ Polyakov model,  we obtain 
    	$ (\Z_N^{[0]})_{\rm M}$ form symmetry in the $PSU(N)$ theory. This process  extends the periodicity of the dual photon field from the weight 
lattice  to root lattice,   $\bm{\sigma}\sim \bm{\sigma}+2\pi \bm{\alpha}_i ,  \; 
 \bm{\alpha}_i  \in \Gamma_r$. The action of the  	$ (\Z_N^{[0]})_{\rm M}$ is to shift $\bm{\sigma} \rightarrow \bm{\sigma}+2\pi \bm{\mu}_i ,  \; 
 \bm{\mu}_i  \in \Gamma_w$, and the  order parameter for this symmetry is monopole operators with magnetic charges in the co-weight lattice,    $e^{i \bm{\mu}_i \cdot \bm{\sigma}(x)}$. In the infrared, 
    	we  obtain a spontaneously broken   $\mathbb Z_N^{[0]}$  0-form symmetry. As a result, there exists $N$ vacua in the thermodynamic limit of the $PSU(N)$ theory distinguished by the vacuum expectation values of the monopole operators  
	\begin{align}
	\langle e^{i \bm{\mu}_i \cdot \bm{\sigma}(x)} \rangle = e^{i  \frac{2 \pi k}{N} },  \qquad k=0,1, \ldots 
	\end{align}
and the infrared limit of $PSU(N)$ Polyakov model is a  $\Z_N$ TQFT.   For dYM, the global symmetry becomes  $(\Z_N^{[0]})_{\rm M}  \times  (\Z_N^{[1]})_{\rm M}$. The realization of this symmetry, say for $p=0$ and varying  $\theta \in [0, 2\pi N)$  subtly depends on the range of theta angle and we will not discuss it here.

\subsection{Non-perturbative mass spectra for Polyakov vs. dYM}
In both Polyakov model and dYM, monopole instantons induce a non-perturbative potential that gaps out all dual photons non-perturbatively. The potential is 
  \begin{align}   
{\rm P:} \qquad    {\cal L}_{m,1}   &=  -    2 \xi     \sum_{i=1}^{N-1}   \,  \cos\left(\bm{\alpha}_i  \cdot \bm{\sigma} \right)  \qquad  \qquad \;\;  {\rm or}   \qquad 
\alpha_i \in  \Delta^0     \cr  
{\rm  dYM:} \qquad  {\cal L}_{m,2}   &=   -  2 \xi    	 \sum_{i=1}^{N}   \,  \cos\left(\bm{\alpha}_i  \cdot \bm{\sigma} + \frac{\theta}{N} \right)    
 \qquad {\rm or}  \qquad  \alpha_i \in \widehat   \Delta^0
\label{mon-roots} 
  \end{align} 
In Polyakov model, the leading saddles are  monopoles in simple root system $\Delta^0= \{ \alpha_1, \ldots, \alpha_{N-1} \} $, 
 while in dYM, monopoles in  extended (affine) simple root system   $\widehat \Delta^0= \Delta^0 \cup \{ \alpha_{N} \}$  are contributing at leading order in semi-classics.  This follows from the fact that the latter theory is locally four dimensional, and the affine root is in the same footing with the monopoles in $\Delta^0$.

To find the dual photon masses, we can expand non-perturbatively induced potential \eqref{mon-roots} to quadratic order   as   $ \xi \sigma^T Q \sigma$. 
  In writing the quadratic potential for  $\bm \sigma$, we use a basis of $N$-component vectors 
$(\sigma_1, \ldots, \sigma_N)$, where one  component  corresponding to  $\frac{1}{\sqrt N} (\sigma_1+ \ldots+ \sigma_N)$  is  redundant, and  
decouples.  In both cases, we  have  $N-1$ physical dual photons only.   
  \begin{align}   
{\rm P:} \qquad    {\cal L}_{m,1}   &=  {\textstyle  \xi}  \sum_{j=1}^{N-1}     \;  ( \sigma_{j} - \sigma_{j+1} )^2   \cr  
{\rm  dYM:} \qquad  {\cal L}_{m,2}   &=  { \textstyle  \xi}   \sum_{j=1}^{N}   \;   \; ( \sigma_{j} - \sigma_{j+1})^2, \qquad \sigma_{N+1} \equiv \sigma_1
  \end{align} 
  For example,  for $N=7$ theories, the $Q$-matrix that needs to be diagonalized is given by:
  \begin{align}
\begin{array}{cc}
Q_{\rm P} = \left[
\begin{array}{ccccccc}
 1 & -1 & 0 & 0 & 0 & 0 & 0 \\
 -1 & 2 & -1 & 0 & 0 & 0 & 0 \\
 0 & -1 & 2 & -1 & 0 & 0 & 0 \\
 0 & 0 & -1 & 2 & -1 & 0 & 0 \\
 0 & 0 & 0 & -1 & 2 & -1 & 0 \\
 0 & 0 & 0 & 0 & -1 & 2 & -1 \\
 0 & 0 & 0 & 0 & 0 & -1 & 1 \\
\end{array}
\right]  \qquad  &  
Q_{\rm dYM} =
\left[
\begin{array}{ccccccc}
 2 & -1 & 0 & 0 & 0 & 0 & -1 \\
 -1 & 2 & -1 & 0 & 0 & 0 & 0 \\
 0 & -1 & 2 & -1 & 0 & 0 & 0 \\
 0 & 0 & -1 & 2 & -1 & 0 & 0 \\
 0 & 0 & 0 & -1 & 2 & -1 & 0 \\
 0 & 0 & 0 & 0 & -1 & 2 & -1 \\
 -1 & 0 & 0 & 0 & 0 & -1 & 2 \\
\end{array}
\right] 
\end{array}
\end{align}
Diagonalizing the mass matrix, we find the non-perturbative mass spectrum in both theories. Indeed, in both cases,  there is a zero eigenvalue  
corresponding  to  unphysical   mode and  is removed from the spectrum.  
Masses of $N-1$ dual photons in Polyakov model are:
 \begin{align} 
 \label{specPol}
 {\rm P:} \qquad  \qquad  m_k^2  = m_\gamma^2 \sin^2  \left(   \frac{ \pi  k }{2N}   \right),  \qquad k=1,  \ldots, N-1
 \end{align}
 while the ones in dYM are given by 
 \begin{align} 
  \label{specYM}
{\rm  dYM:} \;\;  m_{k,q}^2 (\theta)  = 
m_\gamma^2 \sin^2  \left(   \frac{  \pi k  }{N}   \right) \cos \left( \frac{\theta + 2 \pi q}{N} \right),  \qquad k  =1,  \ldots, N-1, 
 \end{align}
 and $ q=0, \ldots, N-1 $ mod $N$ is the branch label. 
Setting $\theta=0, q=0$ for proper comparison,  the difference by a factor of two in the argument of the sine function  is indeed there.  This difference can be  
understood in simple terms, as the  difference of  the normal modes and spectrum of    $N$ coupled   springs  and ball  systems.  
 The  dYM  corresponds to a ring of $N$ coupled   oscillators with periodic identifications at the end 
  and the Polyakov model corresponds to  $N$ coupled  oscillators with open boundary conditions and   balls at the  ends.  The normal modes  and  frequencies  of  coupled oscillator systems are the same as photon eigenstates and masses. 
   \footnote{Note that the distance between min and max of the dual photon  mass square is an instanton factor, $\xi$.  In a natural  abelian large $N$ limit,  
   this implies that $N \rightarrow \infty $  states must fit into   a continuous band, and   the gap in the $SU(N)$ model vanishes as $\frac{1}{N}$ in the large-$N$ limit. This construction admits  an interpretation as an emergent dimension \cite{Cherman:2016jtu}.  } 
         	\begin{figure}[t]
    		\begin{center}
   	\hspace{+5.3cm}			\includegraphics[scale=0.4]{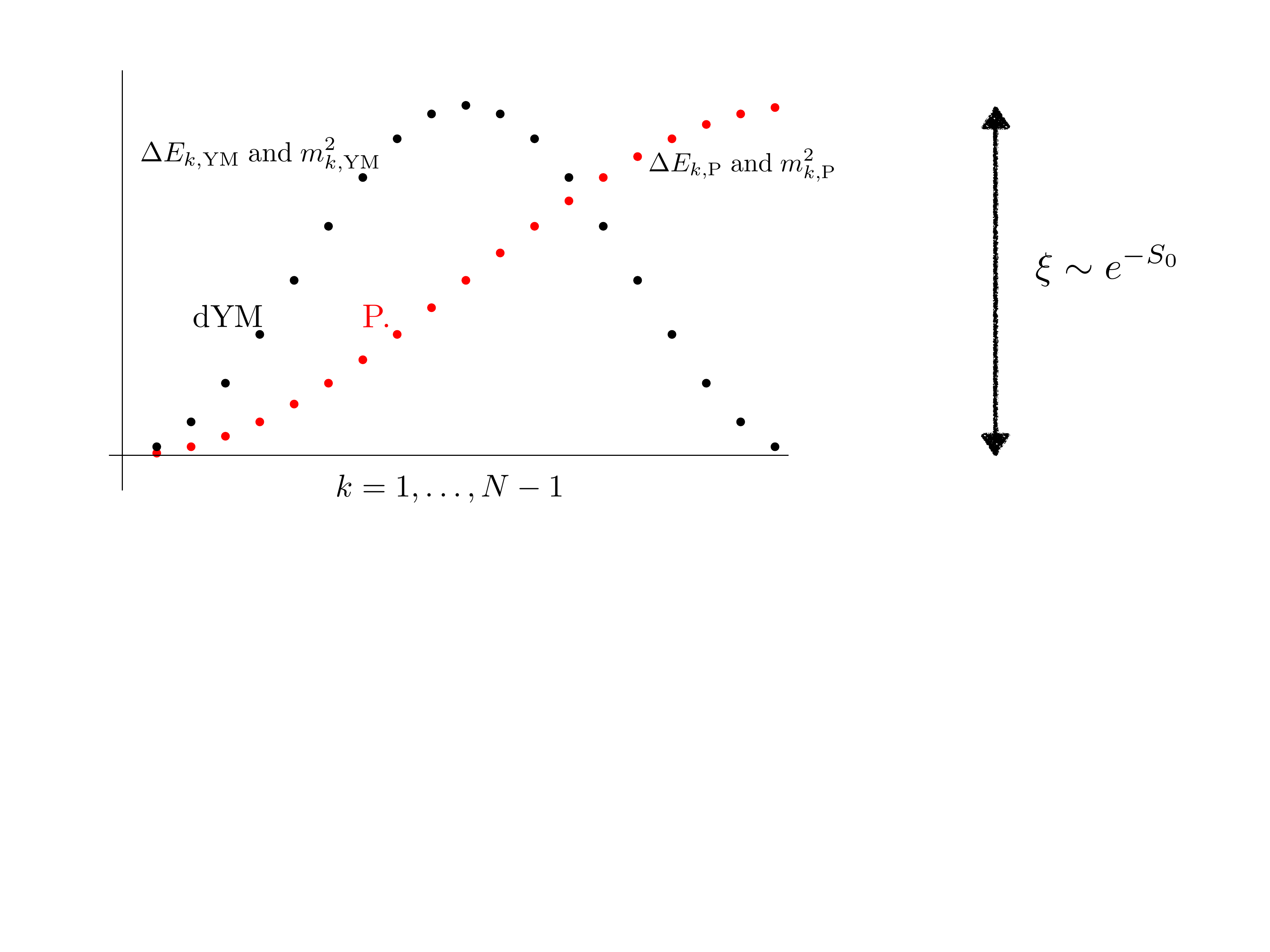}
    		\end{center}
    		\vspace{-5.5cm}
    		\caption{   $\Delta E_k$ is the spectral gaps  in reduced QM.   $m_k^2 $  are mass square for  photons in full QFT,  Polyakov model on $\mathbb R^3$ (red points)  and deformed YM on  $\mathbb R^3 \times S^1$ (black points).  Both have a spectral band width of order $\xi \sim  e^{-S_0}$.  See text.  }
    		\label{MEPdYM}
    	\end{figure}
	
In dYM, the mass spectrum is multi-branched as a function of $\theta$ angle. It is composed of $N$-branches. Note that not all of the  $N$ branches are simultaneously stable.  This is consistent with what is argued in \cite{Witten:1980sp}.  For a given value of $\theta$, approximately  half of the branches  $\sim N/2$  are stable.   Which half of it  is stable depends on the value of $\theta$. 
The mass gap of the system  is given by $k=1, N-1$, and for a given value of $\theta$, it is equal to 
 $m^2_{\rm gap} = {\rm Max}_q m_{1,q}^2 (\theta)$.   See \cite{Aitken:2018mbb} for further details.

\subsection{QM reductions and energy spectra in Born-Oppenheimer approximation} 
Compactifying the Polyakov model on    $\mathbb R \times T^2 $ and dYM theory  on
  $\mathbb R \times T^2 \times S^1$, and inserting  $\ell_{12} =1$  unit of  background  't Hooft flux through $T_2$   
  renders the flux configurations  $  |  {\bm \nu}_{j}   \rangle,  \;  j=1, \ldots N $  degenerate classically.   However, QM description of these two systems are different. dYM has an extra $\Z_N^{0}$ form symmetry  which cyclically permutes the magnetic flux vacua.  The tunnelings in these  two systems connect the vacua as
   \begin{align}
{\rm P:} \qquad  \qquad  &  | {\bm \nu}_1 \rangle \underbrace {\longrightarrow}_{    -{\bm  \alpha}_1}   | {\bm \nu}_2 \rangle   \underbrace {\longrightarrow}_{    -{\bm  \alpha}_2}  
 \cdots 
  \cdots   
  \underbrace {\longrightarrow}_{    -{\bm  \alpha}_{N-1}}    | {\bm \nu}_N \rangle \cr
{\rm  dYM:} \qquad  \;\; {\longrightarrow} &  | {\bm \nu}_1 \rangle \underbrace {\longrightarrow}_{    -{\bm  \alpha}_1}   | {\bm \nu}_2 \rangle   \underbrace {\longrightarrow}_{    -{\bm  \alpha}_2}  
 \cdots 
  \cdots   
    \underbrace {\longrightarrow}_{    -{\bm  \alpha}_{N-1}}    | {\bm \nu}_N \rangle \underbrace {\longrightarrow}_{    -{\bm  \alpha}_N} | {\bm \nu}_1 \rangle
       \label{tunnelilngsQM}
\end{align} 
at leading order in semi-classics.  See Fig.~\ref{PDYM} for a depiction of states and global structure.

   The tight-binding Hamiltonians capturing the tunnelings between the flux vacua $  |  {\bm \nu}_{j}   \rangle $ are  given by:
  \begin{align}  
{\rm P:} \qquad     \widetilde  H_1  &= -  \sum_{j=1}^{N-1}  \xi  \; 
  | {\bm \nu}_{j+1} \rangle \langle  {\bm \nu}_{j} |+ {\rm h.c.}   \cr
{\rm  dYM:} \qquad   \widetilde  H_2  &= -  \sum_{j=1}^{N}  \xi e^{ i  \frac{\theta}{N}} 
  | {\bm \nu}_{j+1} \rangle \langle  {\bm \nu}_{j} |+ {\rm h.c.} 
  \end{align} 
  The  eigenstates of the Hamiltonian  $ \widetilde  H_1 $ for $N=2,3,4$  are given in  Section \ref{sec:tunnel}.  The eigenstates of the $ \widetilde  H_2$ Hamiltonian are much easier to write thanks to the $\Z_N$ cyclic shift symmetry of the corresponding QM. They are given by the discrete Fourier transform of the magnetic flux states:
      \begin{align} 
      |  \Psi_ k\rangle = \frac{1}{\sqrt N} \sum_{j=1}^{N}  e^{i \frac{2 \pi j k }{N}  }      | {\bm \nu}_{j} \rangle,   \qquad k=0, \ldots,  N-1
    \end{align}
The energy eigen-spectra of the low-lying $N$ states are given by
  \begin{align} 
{\rm P:} \qquad   E_k  &= - 2 \xi  \cos \left( \frac{ \pi}{N+1}(k+1) \right),   \qquad k=0, \ldots,  N-1  \cr 
{\rm  dYM:} \qquad  E_k  &= - 2 \xi  \cos \left( \frac{  \theta+ 2 \pi k}{N} \right),   \qquad \qquad \;  k=0, \ldots,  N-1
\label{specth}
  \end{align}  
  The $N$-fold degeneracy is lifted non-perturbatively by instanton effects in Polyakov model and there is no state in the spectrum that remains  degenerate. 
  For dYM,  at generic $\theta$, states are non-degenerate.   The degeneracies occur in a certain pattern for  special values $\theta=0, \pi$ as 
      \begin{align}
     \begin{array}{lll} 
     N-{\rm even} &  \qquad \theta=0 & \qquad { \rm degeneracies:}  \;\; 12\ldots 21 \cr
       N-{\rm even} & \qquad  \theta=\pi & \qquad    { \rm degeneracies:}    \;\;  22\ldots 2 \cr
         N-{\rm odd} &  \qquad \theta=0 & \qquad { \rm degeneracies:}  \;\; 12\ldots 2 \cr
       N-{\rm odd} & \qquad  \theta=\pi & \qquad    { \rm degeneracies:}    \;\;  2\ldots 21
     \end{array}
     \end{align} 
  The exact degeneracy for the $N$ even case  at $\theta=\pi$  is  the realization  of the  mixed anomaly between $\Z_N^{[1]}$  form symmetry and time-reversal  $T$  in QM.   The pattern at odd $N$ is a consequence of global inconsistency, see \cite{Kikuchi:2017pcp}.
      Polyakov model does not possess a mixed-anomaly related to its $\Z_N^{[1]}$ 1-form symmetry and  the lowest-lying $N$ states are non-degenerate. 
   
  The gap between the ground state and higher states in the  energy eigen-spectra of the low-lying $N$ states  in quantum mechanics are given by
  \begin{align} 
& {\rm P:} \qquad   \Delta E_k  =  2 \xi \left(  \cos \left( \frac{ \pi}{N+1} \right) -    \cos \left( \frac{ \pi (k+1)}{N+1} \right) \right),   \qquad k=1, \ldots,  N-1  \cr 
&{\rm  dYM:} \qquad  \Delta E_k (\theta=0 ) =  4 \xi  \sin^2  \left( \frac{  \pi k}{N} \right),   \qquad \qquad \;  k=1, \ldots,  N-1
\label{specth}
  \end{align}      
  $   \Delta E_k (\theta=0 )  $ in dYM quantum mechanics on tiny   $T^2 \times S^1 \times \mathbb R$   is equal to   mass square of the photons $m_k^2 (\theta=0) $   \eqref{specYM}  in full QFT  on $\mathbb R^3 \times S^1$.   
  For  Polyakov model,  the same relation also holds, but  approximately  $\Delta E_k \approx  m_k^2 $, see \eqref{specPol}.
  We find this correspondence quite striking.   In both QFT and QM limits, the width of the non-perturbative band is controlled by 
  monopole-instanton amplitude, and $N$ states fill the interval with the patterns shown in Fig.~\ref{MEPdYM}. 
  
  Note that  the monopole instanton amplitude is controlled by the action
  $S_0 = \frac{4\pi v}{g_3^2}$ in Polyakov model and its QM reduction with 't Hooft flux, and for dYM, since $v = \frac{2\pi}{LN}$, 
  $S_0 = \frac{1}{N} \frac{8 \pi^2}{g_4^2}  $. Despite the fact that the theory is reduced to a tiny $T^2 \times S^1 \times \mathbb R$ 
  in dYM, the ground state properties is controlled by fractional instanton action, $S_0= \frac{1}{N} S_{I_{4d}}$.   
  In dYM, the fractional action is protected by the TQFT coupling. Since we know the resurgence techniques in the QM limit, 
  and since the action of the fractional instantons remain invariant as the $ T^3 \times \mathbb R$ size becomes larger, this construction has direct implications in QFT.  Completeness of this semi-classical basis and the implications of this QM limit to the 
  IR-renormalon problems are discussed in   \cite{Unsal:2021cch}.

  \vspace{0.3cm} 
Note that the max and min of the band are separated by the monopole  instanton factor which is at most  $ \sim  4 \xi$.  Since $\xi$ does not depend on $N$, 
$N$ states must fit into this  interval  in the large-$N$ limit.    This implies that in the large-$N$ limit,  the states in the QM system form a continuous band similar to the full QFT mass spectrum \cite{Cherman:2016jtu}.

\section{Prospects} 
{\bf Cluster expansion vs. Steepest descent:}
An intriguing aspect of our construction is following. Polyakov model, in the semi-classical approximation, maps to a grand canonical ensemble of Coulomb charges. In general, Coulomb gas can be studied via cluster expansion. Cluster expansion involves 
integrals such as (assume for simplicity $q= \pm 1$)
  	\begin{align}\label{PartitionFunction_TwoInstanton-5}
    		Z[q_1, q_2, \ldots, q_N]  &=   \int \prod_{i=1}^{N}  \mrmd^3 {\bf r}_i   \;    \exp \left[ -   \sum_{i < j}  \frac{\pi q_i q_j  }{g^2  | {\bf r}_i -  {\bf r}_j| }  \right]   
    	\end{align} 
and normally,  (if we check  textbooks on statistical physics),  one would  interpret the integral over the configuration space. However, semi-classical analysis instructs us to perform the integrations over the steepest descent cycles of critical points. This does not change the space of integration for repulsive charges, see Fig. \ref{fig:thimbles-1},  but for attractive charges, the steepest descent cycle  lives in the complex domain, see Fig. \ref{fig:thimbles-2}.  For a collection of arbitrary charges, semi-classics treatment requires  complexification of configuration space.  
  As a result,  for example  for $N=2$, we obtain: 
  	\begin{align}\label{PartitionFunction_TwoInstanton-6}
    		Z[+,+]  &= Z[-,-]   = \V (\V + I(g^2) ) \cr 
			Z[+,-]  &= Z[-,+]   = \V (\V -  I_{\pm}(g^2) )  
    	\end{align} 
In the sense of involving  complex and two-fold ambiguous results such as $ I_{\pm}(g^2)$, our construction differs from standard cluster expansion in classical statistical  mechanics.  Perhaps, a better way to treat Coulomb gas in cluster expansion must go through this process of complexification.   It is desirable to understand this aspect at a deeper level.  

Even the minus in front of the second term in \eqref{PartitionFunction_TwoInstanton-6}, which arises naturally from the steepest descent path is physical and important. For example, if we consider ${\cal N}=2$ SYM theory in 3d with $SU(2)$ gauge group, first-order  monopole terms in semi-classics do not contribute to bosonic potential due to index theorem. At second order, there are $ [{\cal M}  \overline{\cal  M}]$ configurations.   The fact that this configurations contributes to the bosonic potential as $V(\tilde \phi) = + \xi^2 e^{-2 \tilde \phi}$  arises from the minus sign in \eqref{PartitionFunction_TwoInstanton-6}. (This leads to the physically correct consequence of positive potential and run-away vacua at infinity \cite{Affleck:1982as}.   If it was not for the minus sign in \eqref{PartitionFunction_TwoInstanton-6}, we would obtain $V(\tilde \phi) = - \xi^2 e^{-2 \phi} <0$ which is negative. This would be  in contradiction with supersymmetry algebra which demands that vacuum energy must be positive semi-definite.

\vspace{0.3cm}
\noindent
{\bf Relation to renormalization group:}
Another aspect is following. Sub-extensive  factors in  \eqref{PartitionFunction_TwoInstanton-6} correspond to contribution of 2-clusters to vacuum energy density.  The 2-clusters are  associated with $+2$ charges, $-2$ charges and dipoles with charges $0=1-1$.   Naturally, there is a sense in which cluster expansion is tied with the renormalization group, since we are integrating over the separations between instantons. If the 2-cluster  is supported 
at separations $r  \lesssim r_b$ for some length scale $r_b$, then  it is meaningful to coarse-grain the system up to length scale $r_b$,  and view the 2-cluster operators as independent operators in an EFT valid at length scales larger than $r_b$.  The 2-cluster operators obtained in this way are called 
bion operators. 

\vspace{0.3cm}
 If we start with microscopic interactions ${\cal L}_{r_m} \supset - \xi (e^{i \sigma} +  e^{-i \sigma})$ where $r_m$ is the short-distance cut-off, the  2-cluster  effects would be of the form  
 \begin{align}
 &- \xi^2 I(g^2)  e^{2 i \sigma (\bf r)}, \qquad -  \xi^2 I(g^2)  e^{-2 i \sigma ((\bf r) )},  \cr
 &  +  \xi^2 I_{\pm}(g^2)   \left( 
  e^{ i \sigma (\bf r +  \bm \epsilon)}  e^{ -i \sigma (\bf r )}   +   c.c \right) \sim   \xi^2 I_{\pm}(g^2)  \cos   (  \bm \epsilon \cdot  \nabla \sigma  ) 
\end{align}
where  $\bm \epsilon$ is the direction of the dipole, and  $|\bm \epsilon| \lesssim r_b$.   Since dipole have a directionality, and they  can come in any direction, we must  average over  the  directionality  to obtain:  
 \begin{align}
 &   I_{\pm}(g^2)   \xi^2    \left(  1-  \frac{1}{6} (\nabla \sigma)^2    + \ldots \right)
\end{align}
In the leading term 
${\rm Re}[ I_{\pm}(g^2)]  \xi^2 \times 1$ clearly contributes to vacuum energy density  (proportional to second virial coefficient) 
and ${\rm Re}[ I_{\pm}(g^2)]    \xi^2 \times  \frac{1}{6} (\nabla \sigma)^2 $ 
 corresponds to renormalization of the kinetic term of the dual photon at second order in semi-classics, and higher-order terms are suppressed by higher powers of derivative.  Both terms are quite physical.  The latter term reminds us of the physical fact that if we were to consider just a gas of dipoles, 
 it would not generate a mass gap, and dipoles would only modify the dielectric constant of the medium.   In fact, by using this line of reasoning, we would immediately prove that gas of dipoles do not induce a mass gap or screen \cite{Glimm:1987ng}.
 So far, it all makes sense.
 
 \vspace{0.3cm}
 But semi-classics also tells us that there is an imaginary ambiguous part in second-order contribution, ${\rm Im}[ I_{\pm}(g^2)]  \xi^2$. As emphasized in the paper, this is not a bug, but a feature. Indeed, according to resurgence, this ambiguity must cancel the ambiguity that arises from the  left/right Borel resummation of perturbation theory, as in \eqref{res11} and \eqref{res12}.  In fact, in the quantum mechanical limit,  this is already proven by resurgence methods.  But here is a subtle issue. 
 If we were to consider coarse-graining in the renormalization group rather than integration over the steepest descent cycle in the semi-classics, 
 naively, we would not obtain the  imaginary ambiguous parts in  $[ I_{\pm}(g^2)]  \xi^2$, but just its real unambiguous part.  A natural question is, 
 where do the imaginary ambiguous parts enter the story in the renormalization group procedure?

\vspace{0.3cm}
\noindent
{\bf Other abelianizing  theories:} Several interesting QFTs undergo dynamical abelianization at long distances and that possess a 1-form center symmetry.  The 4d examples are  ${\cal N}=1$ SYM and QCD(adj)  on small $\mathbb R^3 \times  S^1$ with periodic boundary conditions on fermions on $S^1$,     ${\cal N}=2$  and  ${\cal N}=4$ SYM  on 
$\mathbb R^4$,  and $SU(N) \times SU(N)$ QCD with bifundamental fermions and double-trace deformations on  small $\mathbb R^3 \times  S^1$, and its generalization to chiral quiver theories $SU(N) \times\ldots \times  SU(N)$.   The 3d examples are 
various generalizations of Polyakov models with adjoint fermions  on $\mathbb R^3$, as well as variants as in 4d.  To understand different dynamical consequences, the physical set-up we employ, either on $T^2 \times S^1 \times \mathbb R$  
or $T^2  \times \mathbb R$ with magnetic GNO  flux on $T^2$  (corresponding to a non-trivial 't Hooft flux, but less abstract thanks to abelianization) can be used   to understand different dynamical behaviors in these theories.  
 In all these systems, the monopole tunneling events must be operative, but they  must lead to vastly different  dynamical consequences due to the interplay of  fermionic zero modes and  the existence of extra scalars in extended supersymmetric theories. In some sense, the idea is to bring  long-distance semi-classical calculability with the 't Hooft flux to understand detailed dynamics.  
 We expect this to be a fertile playground to study a diverse set of theories.

    	\section*{Acknowledgements}
We also thank Yuya Tanizaki, Aleksey Cherman and Mendel Nguyen for discussions. M.\"U. acknowledges support from U.S. Department of Energy, Office of Science, Office of Nuclear Physics under Award Number DE-FG02-03ER41260.
C.P is supported by TUBITAK 2214-A Research Fellowship Programme for Ph.D. Students.
	
    	\appendix
    	
\section{Mapping QFT instantons to QM instantons by the method of replicas}
\label{sec:mapping-MI}
In this section, we describe how the monopole in QFT with  magnetic field $\mbfB = \frac{Q}{4\pi r^2}  {\bf \hat r}$ on $\mathbb R^3$ maps to instantons in QM on  small $T^2 \times \mbbR$.  
In QM, we describe the instantons as flux changing events through $T^2$.   
Defining      	$	{\bm \F} (\tau)  = \int_{12}  {\bm B} $, 
the tunneling events correspond to 
\begin{align}
  	\Delta  {\bm \F} =  {\bm \F} (\tau=\infty)- {\bm \F} (\tau=-\infty) = 	 \frac{2 \pi}{g} \alpha, \qquad \alpha \in \Gamma_r^{\vee} .
\end{align}
Below,  we  show that indeed $\bm B$ becomes independent of $(x,y)$ coordinates rather quickly  and   $\bm \Phi(\tau) $ acquires  a standard quantum mechanical instanton profile.   We identify $\tau$  in QM with $z$ coordinate in QFT.  

 \vspace{0.3cm}
Assume $T^2$ is symmetric, with sizes $L_1= L_2= L$.  We take  $L$ much larger than the monopole core size $r_m$ and much smaller  
than characteristic monopoles separation  $d_{\rm mm}$  on $\mathbb R^3$: 
\begin{align}
r_m \ll  L   \ll d_{\rm mm}.
\label{scale-sep}
\end{align}
In this way, we guarantee that generically there exists a single monopole per  $T^2$,  and the change in the local  monopole profile relative to $\mathbb R^3$  is negligible. Furthermore,  in the reduced QM system,  the monopoles form a dilute gas in the  Euclidean time direction  $\mathbb R $. 

 \vspace{0.3cm}
Let us  consider a single monopole on $T^2$,  located at  $(0,0,0)$.  We would like to determine the magnetic field at $(x,y, z)  \in T^2 \times \mathbb R$. We  identify $z$ with the Euclidean time direction.   The easiest way to proceed is to use the method of replicas.  Because of the periodic boundary conditions, we can extend the charges periodically  and form a charge lattice where charges are located at $(m, n)  L \in \Z \times \Z$. (See Fig. \ref{monopolelattice}).
\begin{figure}[h]
	\includegraphics[width=\linewidth]{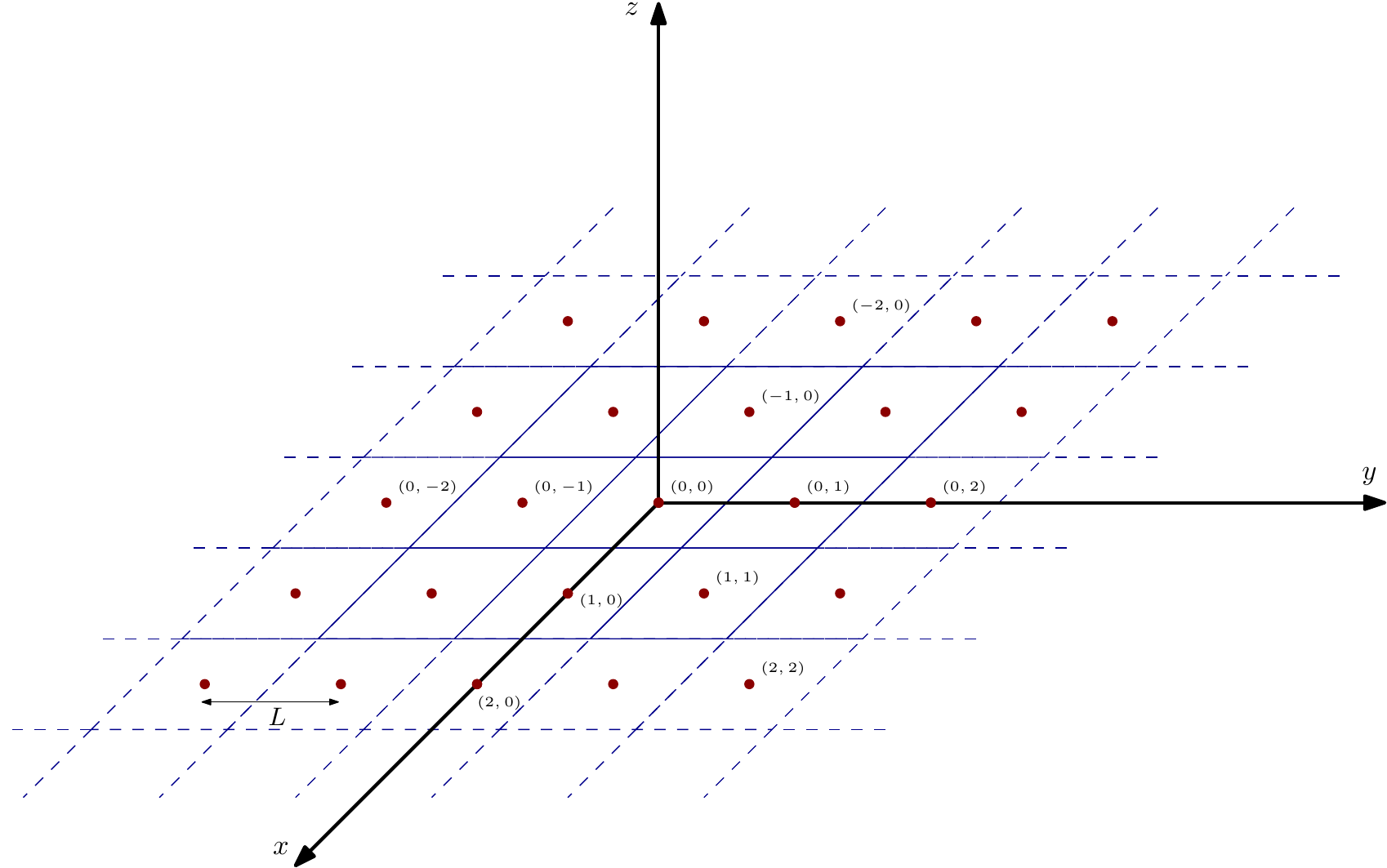}
	\caption{ Each square of size $L\times L$ represents a torus $T^2$ in the compactified $x,y$ dimensions, with a monopole at the center.   Due to periodic boundary conditions, the torus is replicated to form a two dimensional infinite array with monopoles located at $(m,n)L \in \mbbZ \times \mbbZ$.  The field of the  array of monopoles within small $T^2 \times \mathbb R$ is the one of a quantum mechanical instanton. 
	}
	\label{monopolelattice}
\end{figure}
 
 \vspace{0.3cm}
Since the magnetic field can be written as  $\mbfB (x,y,z)  = - \nabla V(x,y,z)$  and it is easier to determine   $V(x,y,z)$ ``the magnetic potential",  let us do so due given  charge distribution. Our goal is to show that the potential  (and magnetic field) becomes independent of the $(x,y)$ coordinate to very good accuracy  for $z \gtrsim  L$. To show this, let us write 
the potential at a point $(x,y,z)$   as a sum over the potential induced by individual  charges:
\begin{equation}
V(x,y,z) = \sum_{m,n} V_{m,n}(x,y,z) = \frac{Q}{4\pi} \sum_{m,n} \frac{1}{
\Big[(x - m L)^2 + (y - n L)^2 + z^2 \Big]^{1/2} }.
\end{equation}
Because of the periodic boundary conditions, the potential is a periodic function in $x, y$. 
\begin{equation}
  V (x+L,y,z) =   V (x,y+L,z) = V(x,y,z)  
\end{equation}
Hence,  we can expand it to  a Fourier series
\begin{equation}
	V(x,y,z) = \sum_{k_x,k_y} V_{k_x,k_y}(z)\, e^{\frac{2\pi i}{L}(k_x x + k_y y)}.
\end{equation}
where $( k_x,k_y) \in \Z \times \Z$ and 
the  Fourier coefficients  are only functions of $z$.  
\begin{align}
	V_{k_x,k_y}(z) &=  \frac{1}{L^2} \int_{0}^{L} \int_{0}^{L}\mrmd x\, \mrmd y\,  V (x,y,z) \, e^{\frac{2\pi i}{L}(k_x x + k_y y)} \cr  
	&=  \frac{1}{L^2} \sum_{m,n}  \int_{Lm }^{L(m+1)} \int_{Ln}^{L(n+1)}\mrmd x\, \mrmd y\, V_{0,0}(x,y,z) \, e^{\frac{2\pi i}{L}(k_x x + k_y y)} \cr
	& = \frac{Q}{   4\pi L^2}  \int_{-\infty}^\infty \int_{-\infty}^\infty \mrmd x\,\mrmd y\, \frac{1}{\sqrt{x^2 + y^2 + z^2}} e^{\frac{2\pi i}{L}(k_x x + k_y y)}
\end{align}
It is convenient to express the last expression in the cylindrical  coordinates. Defining $\bm \r = (x,y) $ and $\bm k = (k_x,k_y)$, we re-write the Fourier coefficients as
\begin{align}
V_{\bm k }  (z)& =  \frac{Q}{4\pi L^2}  \int_0^\infty \mrmd \r \, \r \int_0^{2\pi}\mrmd \t \frac{e^{\frac{2\pi i}{L}(|k|\r \cos\t)}}{\sqrt{\r^2 + z^2}} = \frac{ Q}{2 L^2} \int_0^\infty\mrmd \r  \r \frac{J_0\left( \frac{2\pi|k|\r}{L} \right) }{\sqrt{\r^2 + z^2}}. 
\end{align}  
For $\bm k \neq 0$, we get
\begin{equation}
V_{\bm k \neq 0}(z) = \frac{Q}{2 L^2} \; \Big(\frac{ L}{2\pi |k|}\Big)   \; e^{- \frac{2\pi |k| |z|}{L} }
\end{equation}
Then, the associated magnetic field is
\begin{equation}
B_z^{\bm k\neq 0} = -\frac{\mrmd V_{\bm k\neq 0}}{\mrmd z} =  \frac{Q}{2 L^2}   \mathrm{sign}(z)  e^{-\frac{2\pi |k||z| }{L}}\; ,\quad |k|\neq 0 \label{MagneticField_nonzeromode}
\end{equation}
where ${\rm sign}(z)$ indicates the magnetic field is always in outward direction from the plane of charges, and exponentially decaying rather rapidly. 
For the  zero mode component, $\bm k=0$, we find:
\begin{equation}
B_z^{\bm k=0} (z) = -\frac{\mrmd V_{\bm 0}(z)}{\mrmd z} =   \frac{Q}{2L^2} z  \int  \mrmd \r  \r  \frac{1} { ( \r^2 + z^2)^{3/2}} =  
\frac{Q}{2 L^2}  {\rm sign}(z). \label{MagneticField_zeromode}
\end{equation}
which does not decay at all as expected. 

 \vspace{0.3cm}
For  $\bm k \neq 0$, all the Fourier coefficients   decay  exponentially fast, as in \eqref{MagneticField_nonzeromode}. Even for $|k|=1$ and  at a distance $z= L$, the decay of the mode is $e^{- 2 \pi} \sim 10^{-3}$, and higher modes decay even faster. Therefore, the only mode that matters at distances $|z| \gtrsim L $ is the zero mode of the Fourier decomposition in \eqref{MagneticField_zeromode}.

 \vspace{0.3cm}
This is nothing but the magnetic field due to  uniformly charged infinite plane, with magnetic charge density $Q/L^2$.   The magnetic flux passing through $T^2$ and the change in the magnetic flux between $z \equiv \tau= \pm L$  are given by 
\begin{align}
{\bm \F} (\tau  ) = \left\{ \begin{array}{ll}
 + \frac{Q}{2} & \qquad  \tau > L  \\
-  \frac{Q}{2} & \qquad  \tau < - L  
\end{array}
\right. \qquad  
  	\Delta  {\bm \F} = Q=  	 \frac{2 \pi}{g} {\bm  \alpha} .
\end{align}
These are the asymptotics of an  instanton in quantum mechanics. 

 \vspace{0.3cm}
If we make this construction for the monopole in deformed YM on $T^2 \times  \mathbb  S^1 \times S^1_\beta$, the action of the 
instanton in reduced QM  is  $S_0 = \frac{1}{N} S_{I_{4d}} $, and this is topologically protected 
because  we can turn on two discrete fluxes in  $H^2(T^4, \mathbb Z_N)$  such that $Q= \frac{ (\ell \cdot m)}{N}  = \frac{1}{N}$. 
Since monopole solutions satisfy  is BPS and satisfy appropriate boundary conditions, this shows that $S_0 = \frac{1}{N} S_{I_{4d}} $.  For Polyakov, there is no such topological reason for the action of the monopole to remain the same as we pass from QFT on $\mathbb R^3$ to QM on  $\mathbb R \times T^2$. But the geometric scale separation   \eqref{scale-sep}  suggests that the action of monopole must remain approximately the same as the one on $\mathbb R^3$.

 \vspace{0.3cm}
 \noindent
{\bf  The role of  't Hooft flux  in elementary description:}
Now, we can translate  the crucial role of  't Hooft flux plays to an elementary  language, in terms of parallel plate ``magnetic capacitors." Assume we do not insert a 't Hooft flux but consider an instanton-anti-instanton event. In the above language, this is a simple parallel plate (magnetic) capacitor.   The energy difference between the system with  and without the capacitor  (whose plates are separated by $d$)  is  $\Delta E = \frac{Q^2}{2} d $, stored in between the capacitors.  Since     $\Delta E  >0$ and $\Delta E \rightarrow \infty $ as $ d \rightarrow \infty$,   the  state in between the parallel plates is not the vacuum state.  This implies that  instantons do not survive in the description of vacuum properties in the Polyakov model without discrete flux. 

 \vspace{0.3cm}
Now, let us insert a judiciously chosen background magnetic flux, say  $\Phi = {\bm \mu}_1$. This is similar to inserting parallel plates at infinity    with 
 charge $ \pm {\bm \mu}_1$ at two ends,   and it  causes a finite zero-point energy per unit length.  Now, consider the anti-instanton and instanton   pair, a magnetic capacitor. The flux  in between the capacitors is   $\Phi = {\bm \mu}_1 - {\bm \a}_1$. Now, the energy difference between the system with  and without the capacitor is  $\Delta E =0 $. Since $\Delta E  =0$ and is independent of the separation,   the state in between the parallel plates  has the same energy as the state outside, they are classically degenerate. 
 Therefore, we can make all the tunneling events in QFT survive in a quantum mechanical limit  in the Born-Oppenheimer approximation by using appropriate  magnetic flux background.

\section{Cluster expansion, semi-classics and  sub-extensiveness}\label{Section: ClusterExpansion}
For a gapped system, the partition function in the $\beta \rightarrow \infty$ limit is dominated by the ground state. Therefore, we can write  the partition function in the  $\beta \rightarrow \infty$  limit as  $Z(\beta) \sim e^{- \V  {{\cal E}_0}}$  where    ${{\cal E}_0}$ is the ground state energy density and $\V  \equiv 
 \V_{d+1} = \beta \V_{d}$ is the Euclidean space-time volume. 
Let us write 
\begin{align}
\label{prolif}
e^{- \V  {\cal E}_0} &\sim  e^{  \V \big(  [\C_1] + [\C_2] +  [\C_3] +   [\C_4] +  \ldots  \big) }
\end{align}
where $ [ {\boldsymbol{\cdot}} ] $  denotes class of  saddle contributions organized according to their action  $S\sim n S_0 $ or fugacities $\xi \sim e^{-n S_0}$.   For example, 
\begin{align}
\label{clustering}
[\C_1] & \equiv \Big \{ {\cal M}_{\alpha_i},  \;  \overline {\cal M}_{\a_i}, \;   \; i \in [1, N-1] \Big \}  \cr 
 [\C_2] &=  \Big\{  [ {\cal M}_{\alpha_i} \overline {\cal M}_{\a_{i+1}}],    [ \overline {\cal M}_{\alpha_i}  {\cal M}_{\a_{i+1}}],  \; i \in [1, N-2],  \;\;\;   [{\cal M}_{\alpha_i} \overline {\cal M}_{\a_{i}}]_{\pm},  \; i \in [1, N-1]  \Big \}  \qquad 
\end{align}
$ [\C_1]$  denotes first order effects in semi-classics. These   are  regular  saddles.  
$[\C_2]$ are second order effects  in semi-classics. Their relation to saddles in  $ [\C_1]$    is  following. Two configuration in $[\C_1]$, if they interact,  are not genuine saddles at any finite separation, but  they become saddles at  infinite separation. The integral over the steepest descent cycle produce two-effects:  one can be interpreted as
$\frac{1}{2!}\V^2 [\C_1]^2$, non-interacting two-instanton contribution and the other is $\V [\C_2]$, a 2-cluster contribution.  The proliferation of these two type contributions exponentiate to produce  the first two terms in \eqref{prolif}.  Magnetic and neutral bions are the elements of 2-cluster 
configurations.\footnote{\label{footnote: BPS_limit}In the BPS limit $ \lambda \rightarrow 0$, $ {\cal M}_{\alpha_i} $does not interact with  $ {\cal M}_{\alpha_{i+1}} $ as the interaction due to dual photon exchange is cancelled exactly by the scalar exchange.  But there are genuine BPS  solutions with magnetic charge 
$\alpha_i + \alpha_{i+1}$ with action $2S_0$ that needs to be included. These are $ {\cal M}_{\alpha_i + \alpha_{i+1}}$ that generate only {\it quantitative} differences, and to simplify the discussion,  we will not include them below.  But it is necessary to include the second order effects in 
 $[\C_2] $ because they generate qualitative new effects, such as ambiguities, and they naturally arise as part of the critical point at infinity of configurations in $[\C_1] $, while  $ {\cal M}_{\alpha_i + \alpha_{i+1}}$ do not arise in such manner.} 
The  semi-classical expansion is based on class  $ [\C_1]  $ and  $k$-clusters $[\C_k]$ (which arise naturally from critical points at infinity.)

Let us write the partition function as 
\begin{align}
e^{- \V  {\cal E}_0} &\sim  e^{  \V \big(  [\C_1] + [\C_2]+  [\C_3] +   [\C_4] + \ldots  \big)  }\cr
& =  \prod_{\cal T}   \sum_{n_{\cal T}=0}^{\infty}    \frac{1}{n_{\cal T}! } \left[  \V {\cal T}   \right]^{n_{\cal T}} \cr
&= 
\sum_{   n_1= 0}^{\infty}   \frac{1}{n_1 ! }   (\V  [\C_1])^{n_1}  \sum_{   n_2= 0}^{\infty}   \frac{1}{n_2 ! }   (\V  [\C_2] )^{n_2} 
\sum_{   n_3= 0}^{\infty}   \frac{1}{n_3 ! }   (\V  [\C_3] )^{n_3}    \ldots \qquad \qquad    \cr
&=    \sum_{   n_1, n_2,  n_3, \ldots}   \frac{1}{n_1!  n_2!  n_3! \ldots}  \V^{n_1+ n_2+ \ldots}   ([\C_1])^{n_1}  ([\C_2] )^{n_2}  ([\C_3] )^{n_3} \ldots 
\label{EFT2}
\end{align} 
where  $[\C_k]$ is the sum over the configurations that are in the $k$-cluster, examples of which are shown in \eqref{clustering}. 
 For example,   
 $[\C_1]  = \sum_{i=1}^{N-1} ({\cal M}_{\alpha_i} +  \overline {\cal M}_{\a_i}) $
 etc.    So, the cluster expansion gets crowded rather quickly.

To highlight  our main point, we will consider a very simple system whose merits carry over to the current problem 
without cluttering the expression more than necessary. Consider for example, a simple system like particle on a circle in the presence of a potential 
$V(q)= -\cos(q),  \; q \in [0, 2\pi]$. There is a unique minimum on the circle and there is an instanton and anti-instanton.  To clarify the structure of the cluster expansion as much as possible, let us only keep track of  instanton in the sum.  (Anti-instantons can be incorporated 
easily.)  
Let us denote  the $k$-clusters as 
\begin{align}
[\C_k]  =  a_k(g^2)  \xi^k,  \qquad \qquad  a_1(g^2) =1, \qquad a_k (g^2) =P_{k-1} (\log(1/g^2)) \label{k_clusters}
\end{align} 
and use $\V= \beta$.  Now, we  
we can reorganize the partition function in the   $\V \rightarrow \infty$ limit as an expansion in $\xi$ rather than an expansion in  $\V$. 
This will illuminate some aspects of the semi-classics. 
\begin{align}
e^{- \V  {\cal E}_0} &=    \sum_{   n_1, n_2, n_3, \ldots}   \frac{1}{n_1!  n_2! n_3! \ldots}  \V^{n_1+ n_2+ n_3+ \ldots}   a_1^{n_1}
a_2^{n_2}  a_3^{n_3}   \ldots  \xi^{n_1+ 2n_2+ 3n_3+  \ldots}   \cr 
& = 1 \cr
& + \xi^1 \left( \V \frac{a_1}{1!}   \right)    \cr
&+  \xi^2 \left( \V^2  \frac{(a_1)^2}{2!} +  \V  \frac{(a_2)}{1!}     \right)  \cr
&+  \xi^3 \left( \V^3  \frac{(a_1)^3}{3!} +  \V^2  \frac{(a_1)}{1!} \frac{(a_2)}{1!}  +   \V  \frac{(a_3) }{1! }   \right)   \cr  
&+  \xi^4 \left( \V^4  \frac{(a_1)^4}{4!} +  \V^3  \frac{(a_1)^2}{2!} \frac{(a_2)}{1!}  +  \V^2  \bigg( \frac{(a_1)}{1!} \frac{(a_3)}{1!} +  \frac{(a_2)^2}{2!}   \bigg)  + 
\V  \frac{(a_4) }{1! }   \right)  \cr
&+  \xi^5 \left( \V^5  \frac{(a_1)^5}{5!} +  \V^4  \frac{(a_1)^3}{3!} \frac{(a_2)}{1!}  +  \V^3  \bigg( \frac{(a_1)^2}{2!} \frac{(a_3)}{1!} +  \frac{(a_1)^1}{1!}   \frac{(a_2)^2}{2!}   \bigg)  +   \V^2  \frac{(a_1) }{1! } \frac{(a_4) }{1! } 
+   \V  \frac{(a_5) }{1! }   \right)    \cr
&+ \ldots 
\label{EFT2}
\end{align}
This expansion require some comments: 
\begin{itemize}  
	\item    Consider order $\xi^n$.  The maximally extensive   part  in $\V$ is a non-interacting gas  of single instantons.

	\item  Let us first describe the meaning of sub-extensive part  for $n=2$.   In \eqref {EFT2} second line,  the overall factor of $\V$ is sourced by integration over the center of action of the two instantons.  The integral over  one QZM direction in    \eqref{EFT2}  yields  $\V  \frac{1}{2!} +    \frac{(a_2)}{1!} $,  a part extensive with volume, and a finite part. The combination of zero mode and quasi-zero mode integration  yields $\xi^2$ term in
\eqref{EFT2}.  Sub-extensive pieces include effects of correlated events, in this case a 2-cluster.   $a_2 (g)$ is a first order polynomial in $\log( 1/g^2)$.  In general,  $a_n (g^2)$ is an $(n-1)^{\rm th}$ order polynomial in $\log( 1/g^2)$. 
	%
	%
	%
	\item At order $\xi^n$,  the terms are  of the form $\V^k    \left(\log( 1/g^2) \right)^{n-k}$.    $\V^n$ part is the non-interacting gas. The last term   
	$\V^{1}  \left(\log( 1/g^2) \right)^{n-1}$ is the contribution of $(n-1)$-cluster, and overall  $\V$ is  the integral over the center position of 
	the cluster. 
	
	\item Finally, and of course quite importantly, quasi-zero mode integration yields $\log( 1/g^2)$ type terms for repulsive interactions and 
	two-fold ambiguous $\log( 1/g^2) \pm i \pi$ for the attractive interactions. As a result, the \eqref{EFT2}   will also  have   two-fold ambiguous terms once anti-instantons are included in the expansion. 
\end{itemize}

 If we apply the same strategy to the Polyakov model, the resulting expressions at second order in semi-classics takes the form
 \begin{align}
\label{clustering2}
& \frac{\V^2}{2!}    [\C_1]^2 +  \frac{\V}{1!}  [\C_2]  \cr
& = \frac{\V^2}{2!}   \Big( \sum_{i =1}^{N-1}  ( [{\cal M}_{\alpha_i}]  + [ \overline {\cal M}_{\a_i}])  \Big)^2 
+  \frac{\V^1}{1!} 
  \sum_{i =1}^{N-2}   \Big( [ {\cal M}_{\alpha_i} \overline {\cal M}_{\a_{i+1}}] +     [ \overline {\cal M}_{\alpha_i}  {\cal M}_{\a_{i+1}}]  \Big)  \cr
& +  \frac{\V^1}{1!} 
  \Big(  \sum_{i =1}^{N-1} \Big(   [ {\cal M}_{\alpha_i} \overline {\cal M}_{\a_{i}}]_{\pm} +     [ \overline {\cal M}_{\alpha_i}  {\cal M}_{\a_{i}}]_{\pm}  \Big) 
\end{align}
which have terms with three different  characteristics. 
\begin{itemize}
\item  Terms which do not have a 2-cluster contribution, eg. $ \frac{\V^2}{2!} [{\cal M}_{\alpha_i}]^2 $ because  ${\cal M}_{\alpha_i}$  does not interact with ${\cal M}_{\alpha_i}$. As such, there is no $[{\cal M}_{\alpha_i} {\cal M}_{\alpha_i}]$ in   $[\C_2]$. 

 \item  Terms which do  have an unambiguous  2-cluster contribution,  e.g.
 \begin{align}\label{2Cluster_Repulsive}
  \V^2 [{\cal M}_{\alpha_i}]  [\overline {\cal M}_{\a_{i+1}}] +  \V   [ {\cal M}_{\alpha_i} \overline {\cal M}_{\a_{i+1}}]  =  \xi^2  ( \V^2 + \V a_2 (g^2)  )    
 \end{align}
 because some instanton and anti-instantons  in  $[\C_1]$ do interact repulsively with each other, and these are contribution of critical points at infinity of such configurations,  integrated over the steepest descent cycle ${\cal J}_1$.  

  \item  Terms which do  have a two-fold ambiguous  2-cluster contribution,  e.g.
   \begin{align}\label{2Cluster_Attractive}
    \V^2 [{\cal M}_{\alpha_i}]  [\overline {\cal M}_{\a_{i}}] +  \V   [ {\cal M}_{\alpha_i} \overline {\cal M}_{\a_{i}}]_{\pm}  =  \xi^2  ( \V^2 + \V a_2 (g^2 e^{ \pm i \pi} )  )  
    \end{align} 
     because some instanton and anti-instantons  in  $[\C_1]$ do interact attractively with each other, and these are contribution of critical points at infinity of such configurations,  integrated over the steepest descent cycles ${\cal J}_2^{\pm}$. These  cycles are  inevitably two-fold ambiguous as described below.  
\end{itemize}

\section{QZM integral by integration by parts} \label{OtherRegularizations}

{\bf Repulsive interaction:} Another way to get the  quasi-zero mode  integral is to first introduce a hard cut-off  $R$  and then, using the integration by parts technique successively. 
Define 
\begin{align}
J(g^2) &= 4\pi v^{-3}\, \lim_{R \rightarrow \infty}  \int_{0}^{R}\mrmd  r\,  r^2 e^{-  \frac{ 2 \pi |\a_i\cdot \a_j|}{g^2  r} }  \cr
 &= 4\pi v^{-3}\, \left(\frac{ 2 \pi |\a_i\cdot \a_j|}{g^2}\right)^3 \int_{0}^{\tilde R } \mrmd \tilde r\, \tilde r^2 e^{- 1/\tilde r }, \qquad \tilde R  =  \frac{g^2}{2 \pi |\a_i\cdot \a_j|}{ R}
\end{align}
The radial integral is well-behaved at $\tilde r=0$ but diverges as $ \tilde r\rightarrow \infty$.  We introduce a hard cutoff at $ \tilde  r= \tilde R$. 
Applying integration by parts repeatedly,   we  give a precise meaning to divergent parts and finite parts.  The divergent part is the volume of the system, and the finite part will be identified with the second virial expansion  parameter, see eg. \cite{kardar}. We can proceed with the repeated application of integration by parts, a procedure which eventually ends up with a finite integration: 
\begin{align} \label{manyint}
{\textstyle \frac{v^3}{4\pi} \left(\frac{\pi |\a_i\cdot \a_j|}{g^2}\right)^{-3}  }J(g^2)  &=   \left( \frac{ \tilde  R^3}{3} e^{-1/ \tilde  R} - \frac{1 }{3}\int_{0}^{\tilde R}\mrmd \tilde r\,\tilde r e^{-1/\tilde r} \right)  \cr
&= \Bigg[ \left(\frac{ \tilde  R^3}{3} - \frac{ \tilde  R^2}{6}  \right) e^{-1/\tilde R} + \frac{1}{6}\int_{0}^{\tilde R}  \mrmd \tilde r  e^{-1/\tilde r} \Bigg]\cr
&= \Bigg[ \left(\frac{ \tilde  R^3}{3} - \frac{ \tilde  R^2}{6} + \frac{ \tilde  R}{6} \right) e^{-1/\tilde R}  - \frac{1}{6}\int_{0}^{\tilde R}  \mrmd \tilde r \frac{ 1}{\tilde r} e^{-1/\tilde r} \Bigg] \cr
&= \Bigg[ \left(\frac{ \tilde  R^3}{3} - \frac{ \tilde  R^2}{6} + \frac{ \tilde  R}{6} - \frac{ \ln \tilde  R}{6} \right) e^{-1/\tilde R} + \frac{1}{6}\int_{0}^{\tilde R}  \mrmd \tilde r \frac{ \log \tilde r }{\tilde r^2} e^{-1/\tilde r} \Bigg] \qquad  \qquad  \qquad  \qquad 
\end{align}
Now, we carefully look at the behaviors of these expressions as $ \tilde R\rightarrow \infty$. In this limit, the last integral is finite, and it  is just the Euler–Mascheroni constant $\g$. The extensive  term can be handled by expanding it around $ \tilde R=\infty$:
\begin{align}
\left(\frac{  \tilde R^3}{3} - \frac{  \tilde R^2}{6} + \frac{  \tilde R}{6} - \frac{ \ln   \tilde R}{6} \right)e^{-1/  \tilde R} & \simeq 
\frac{ \tilde R^3}{3}-\frac{ \tilde R^2}{2}+\frac{ \tilde R}{2} - \frac{11}{36} - \frac{\ln  \tilde R}{6} + O\left(\frac{1}{ \tilde R}\right) 
\label{limit}
\end{align} 
Identifying $ \tilde R = 1/ \delta$,  we observe that this produces the expansion  of incomplete $\Gamma$ function  given in $\eqref{IncompleteGamma_Expanison} $.   As a result, we obtain:
\begin{align}
J_2(g^2) & =  \V  +  I(g^2).
\label{J2-2}
\end{align}
in agreement with \eqref{radialIntegral_cutoffResult3}. 
We have few other comments on this result:
\begin{itemize} 
	\item In  cluster expansion in statistical mechanics, there is a standard trick,  subtracting and adding unity to the integrand:  
	\begin{align} 
	\int \mrmd^3r\,  e^{- V_{\mathrm{int}}(r)} = 	\int \mrmd^3r\, (e^{- V_{\mathrm{int}}(r)}  -1 +1) = 
	\V + \int \mrmd^3r\, (e^{- V_{\mathrm{int}}(r)}  -1)  
	\label{formal}
	\end{align}
	where   $(e^{- V_{\mathrm{int}}(r)}  -1)$  is called Mayer-$f$ function.  But   \eqref{formal} is still formal for Coulomb gas, since the integral is still extensive with volume as $\V^{2/3}$. We believe the procedure described in \eqref{manyint}  and  the limit taken carefully in \eqref{limit}  and also around \eqref{radialIntegral_cutoffResult} are better ways to give meaning to this integral.     $ I(g^2)  $ can be identified  with the second virial expansion  parameter.

	\item  One  of the reasons for writing all steps of the integration \eqref{manyint}  is to point  out  an interesting connection between QFT and QM.  The integral in the third line of 	   \eqref{manyint} $\int  \mrmd \tilde r \frac{ 1}{\tilde r} e^{-1/\tilde r} $  is actually the standard instanton-instanton  interaction term in quantum mechanics.  	  To see this, first, let us restore the coupling  by writing   $ \tilde r  = \frac{g^2}{ 2 \pi |\a_i\cdot \a_j|}{ r} \equiv 
	\frac{g^2}{A}{ r}$ and use change of variables  
	$r= e^{\tau}   $.  Then, the integral becomes 
	\begin{align}
	\int  \mrmd  r  \; \frac{ 1}{  r} \; e^{- \frac{A}{g^2} \frac{1}{r}   }  = \int d \tau  \; e^{ -\frac{A}{g^2}	e^{-\tau} }
	\label{QZM}
	\end{align}
	where $\tau$ is the quasi-zero mode parameter and  $ V_{\rm int} (\tau) = \frac{A}{g^2}	e^{-\tau}$ is the classical interaction between instantons. 
	The interaction is repulsive for $A >0$ and attractive for $A<0$. This type of  integrals  appears  naturally in   QM and  treated   by using the critical point at infinity and Lefschetz thimble integration in  \cite{Behtash:2018voa}.   
	
	\item 	 In QM, the quasi-zero mode integral  is also divergent, with an extensive part $(\beta^1)$ and a finite part  $(\beta^0)$. 
	If we  compactify Euclidean time  $\mathbb R$ to $S^1_\beta$,   at second order in the semi-classical expansion, we obtain  $\frac{\xi^2}{2!} \int_{\Gamma} d\tau_1 d\tau_2   \;  e^{- V_{12} } $, where integration over 
	the exact zero mode $ \frac{1}{2}(\tau_1+ \tau_2)$ produce a factor of $\beta$ and the integral over $\tau_1- \tau_2 = \tau$  is  the QZM integral. 
	 The QZM integral yields  
	\begin{align}
	\int d \tau  \; e^{ -\frac{A}{g^2}	e^{-\tau} }  
	\mapsto  \left\{ \begin{array} {l l }
	(\beta + I_{\rm qm}(g^2))=  \beta +  \log(\frac{A}{g^2} ) + \gamma  &  \qquad A>0\cr \cr
	(\beta + I_{\rm qm}(g^2 e^{\pm i  \pi})=  \beta +  \log(\frac{A}{g^2}) + \gamma  \pm i \pi  & \qquad A<0
	\end{array}  \right.
	\label{QZM2}
	\end{align}
	The QZM integrals in QFT and QM are related in a certain way: 
	\begin{align}
	J_{2, \rm qft} (g^2) & =  \V  +   	4 \pi \left(\frac{2 \pi |\a_i\cdot\a_j|}{g^2}\right)^{3}  \frac{1}{6}  \left(  I_{\rm qm} (g^2) -   \frac{11}{6}   \right)  
	\label{betweendim}
	\end{align}
	where $ I_{\rm qm} (g^2) $ is given in \eqref{QZM2}.  
	
	Note that \eqref{betweendim} connects a  3d calculation to a 1d calculation.  It indicates that cluster expansion in the QFT and QM limit are not independent.  Moreover,   if there is an ambiguity in  $I_{\rm qm} (g^2)$, it indicates that there will be an ambiguity in  $	J_{2, \rm qft} (g^2) $. 
\end{itemize} 

\noindent
{\bf Attractive interaction:}  
The case where the interaction is attractive can be obtained from \eqref{J2-2} by using analytic continuation. The result is 
\begin{align}
J(g^2 e^{\pm i \pi} ) =    \V - I(g^2)  \pm i  \frac{2 \pi^2}{3}  \left(\frac{2 \pi |\a_i\cdot\a_j|}{g^2}\right)^{3}.
\end{align}
Clearly, there is a two-fold ambiguity in the attractive case,  corresponding to   $[ {\cal M}_{\alpha_i} \overline {\cal M}_{\a_{i}}]_{\pm} $ amplitudes. Even after the ambiguity cancels via resurgence relations, note that there is an over-all phase difference between $ \Re[ {\cal M}_{\alpha_i} \overline {\cal M}_{\a_{i}}]_{\pm} $   configuration and  $ [ {\cal M}_{\alpha_i} \overline {\cal M}_{\a_{i+1}}] $. i.e, 
\begin{align}
{\rm Arg}[ \Re[ {\cal M}_{\alpha_i} \overline {\cal M}_{\a_{i}}]_{\pm} ] =  {\rm Arg}  [ {\cal M}_{\alpha_i} \overline {\cal M}_{\a_{i+1}}]  + \pi
\end{align} 
This subtle phenomenon (the fact that these two configurations contribute oppositely to vacuum energy density) appeared earlier in the context of 
${\cal N}=1$ SYM and is called ``hidden topological angle" \cite{Behtash:2015kna}. It is a phase that arises from the difference of the quasi-zero mode integration cycles  
(thimbles) for the repulsive vs. attractive interactions.

	\section{Lefschetz thimbles for Coulomb interaction} \label{Section: Thimbles_CoulombGas}
	
	Consider an  exponential integral  of the form $\int_{\Gamma} dz \;  e^{- \frac{1}{g^2} f(z) }  h(z) $ where $|g^2| \ll 1$. In steepest descent formulation, we first determine the critical points of the action, by setting  $ \frac{ df}{dz}= 0  $, which is satisfied at  $ z= \{ z_1, \ldots, z_N \}$  where $N$ is the number of critical points.  Critical points may be either at finite values of $z$ or $z=\infty$. The latter is called critical point at infinity and its discussion is slightly more subtle.  
	
	Associated with each critical point, one can determine a unique  steepest descent cycle ${\cal J}_i$  (called Lefschetz thimble in higher dimensions) on which $e^{- \frac{1}{g^2}  f(z)}$  is ever decreasing 
	and $e^{- \frac{1}{g^2}  f(z)} \rightarrow 0 $ on certain  wedges  (that we can be called good wedges) in $\mathbb C$.    If $f(z)$ is a polynomial, 
	the cycle ${\cal J}_i$ starts   in some good wedge and ends in some other good wedge as  $ |z | \rightarrow \infty$ 
	and it passes through $z_i$.  (A lucid explanation of this can be found in Witten's work \cite{Witten:2010cx}.)  
	The steepest ascent cycle ${\cal K}_i$  is the cycle on which  $e^{- \frac{1}{g^2}  f(z)}$   is ever-increasing, and it starts and ends at  complementary  wedges  (bad domains), where exponential blows up.  
	If $f(z)$ has a pole, the pole can also serve as an ending point of the thimble.  For example, if  $f(z)$ is a  doubly periodic function on $\mathbb C$, the only place that steepest descent cycle  ${\cal J}_i$ can start and end are the poles as discussed in \cite{Basar:2013eka}.
	

	In our current  example, $f(z) =  \frac{1}{g^2} \frac{1}{z}$. ${\rm Arg}(g^2) = 0$  is for repulsive interactions and  ${\rm Arg}(g^2) = \pi \pm \epsilon $  is for   attractive interactions. In both cases, the critical point is at infinity, and there is a pole at $z=0$.  By using various regularizations, we can  move the critical point to some $R_{*}$, and let it go to infinity as the regulator is removed. Even with the hard-cutoff imposed  in \eqref{manyint}, although $f'(z)$ does not vanish anywhere, it can be made arbitrarily small at the boundary $z=\tilde R$, and the boundary of the integral behaves as a pseudo-critical point, which becomes genuine one as $\tilde R$ tends  to infinity. 
	
	\begin{figure}[h]
		\centering
		\includegraphics[width=0.47\textwidth]{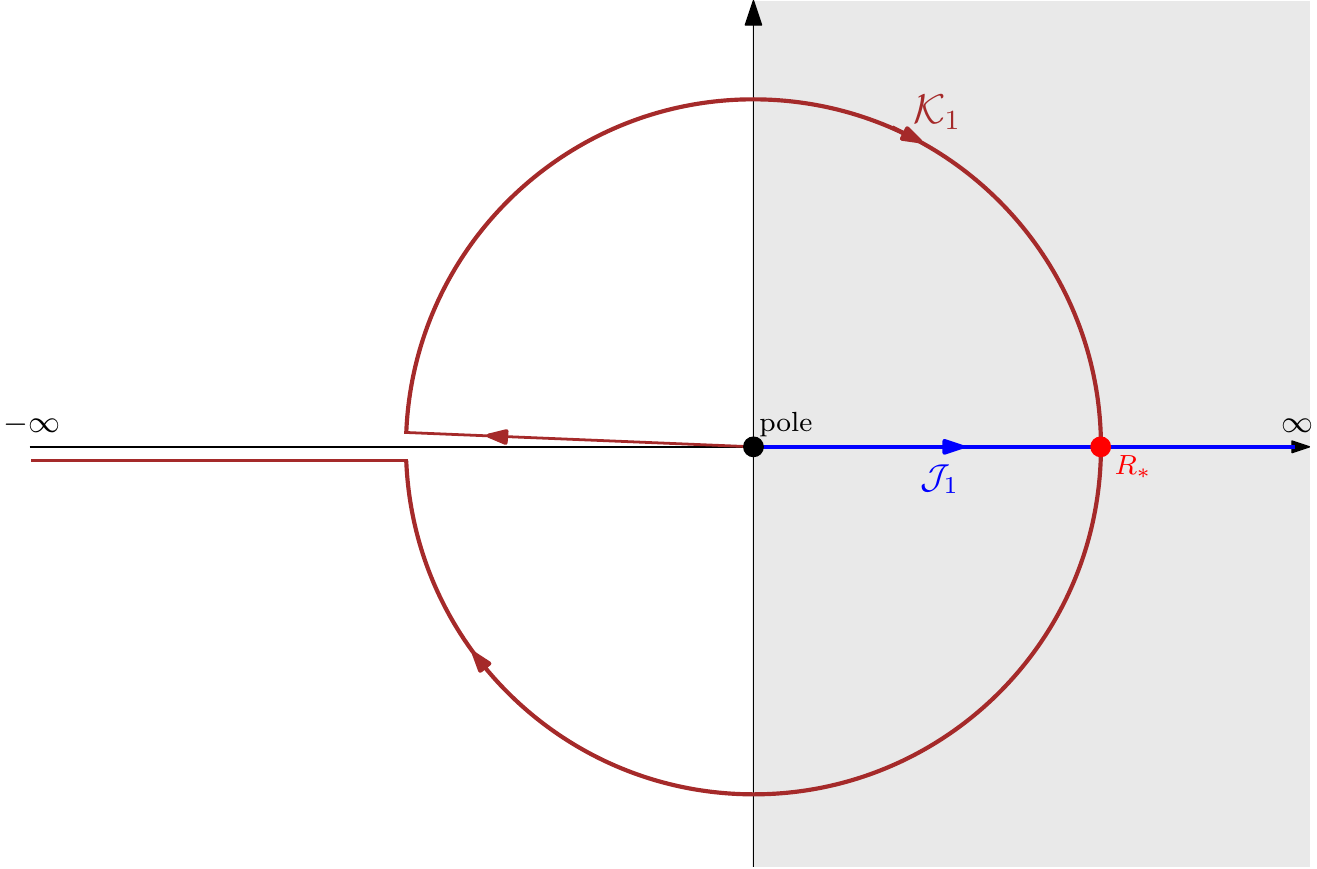}
		\caption{For repulsive interaction,  ${\cal J}_1$  is the steepest descent cycle.   The critical point moves to  infinity as the the  regulator   is removed.  
		${\cal J}_1$  extends from the pole  to   $+ \infty$.    }
		\label{fig:thimbles-1}
	\end{figure}

	\paragraph*{Repulsive interaction:} For the case ${\rm Arg}(g^2) = 0$, the ${\cal J}_1$ cycle is $[0, + \infty)$ as shown in Fig.\eqref{fig:thimbles-1}.  In this case, 
	\begin{align}
		\int_{{\cal J}_1 (0)}  dz  e^{- \frac{1}{g^2} f(z) }  h(z) = {\cal V } + I(g^2) 
	\end{align} 
	Note that ${\cal J}_1$ leaves the  critical point at infinity along the real direction, because it is the direction at which $e^{- \frac{1}{g^2}  f(z)}$ drops fastest. It also enters to the pole in the   ${\rm Arg}(z)=0$ direction, where the approach to zero is fastest.    
	The steepest descent cycle  changes smoothly as we change ${\rm Arg}(g^2) $,  so long as the change in ${\rm Arg}(g^2) $  does not lead to   
	crossing of   a Stokes line.  In the current problem, Stokes line is  at $\t = {\rm Arg}(g^2)=\pi$ which, as we know from the analytical continuation discussion in Section.\ref{Section: CriticalPointsAtInfinty}, corresponds to the attractive interaction.

	\paragraph*{Attractive Interaction:}  First of all, notice that we can move from the repulsive interactions to attractive interactions   by   
	dialing $\theta= {\rm Arg}(g^2) $, and in fact, ${\cal J}_1 (\theta = \pi \pm \epsilon) ={\cal J}_2  (\theta =0 \pm \epsilon)  $ where ${\cal J}_2 $ are the steepest descent cycles for the attractive interaction.   Recall that at $\theta=0$, the  ${\cal J}_1 (\theta)=0$ cycles leaves the origin in $\arg(z)= 0$ direction and enters to infinity at $\arg(z)=0$ direction. This guarantees that the integrand is well-behaved on the thimble in the vicinity of  $z=0$. For $\theta=\pi$, the  ${\cal J}_1 (\theta = \pi \pm \epsilon)$  leaves  the pole at $z=0$ in the  $\arg(z)=\pi \pm \epsilon $ (for convergence of the integral as $|z| \rightarrow 0$) , however, eventually makes a clock-wise or counter-clockwise circle and then tends to $z=-\infty$ as shown in Fig. \eqref{fig:thimbles-2}.  This guarantees that the integral along the thimble  is well-behaved. 
	
	\begin{figure}[h]
		\centering
		\includegraphics[width=0.47\textwidth]{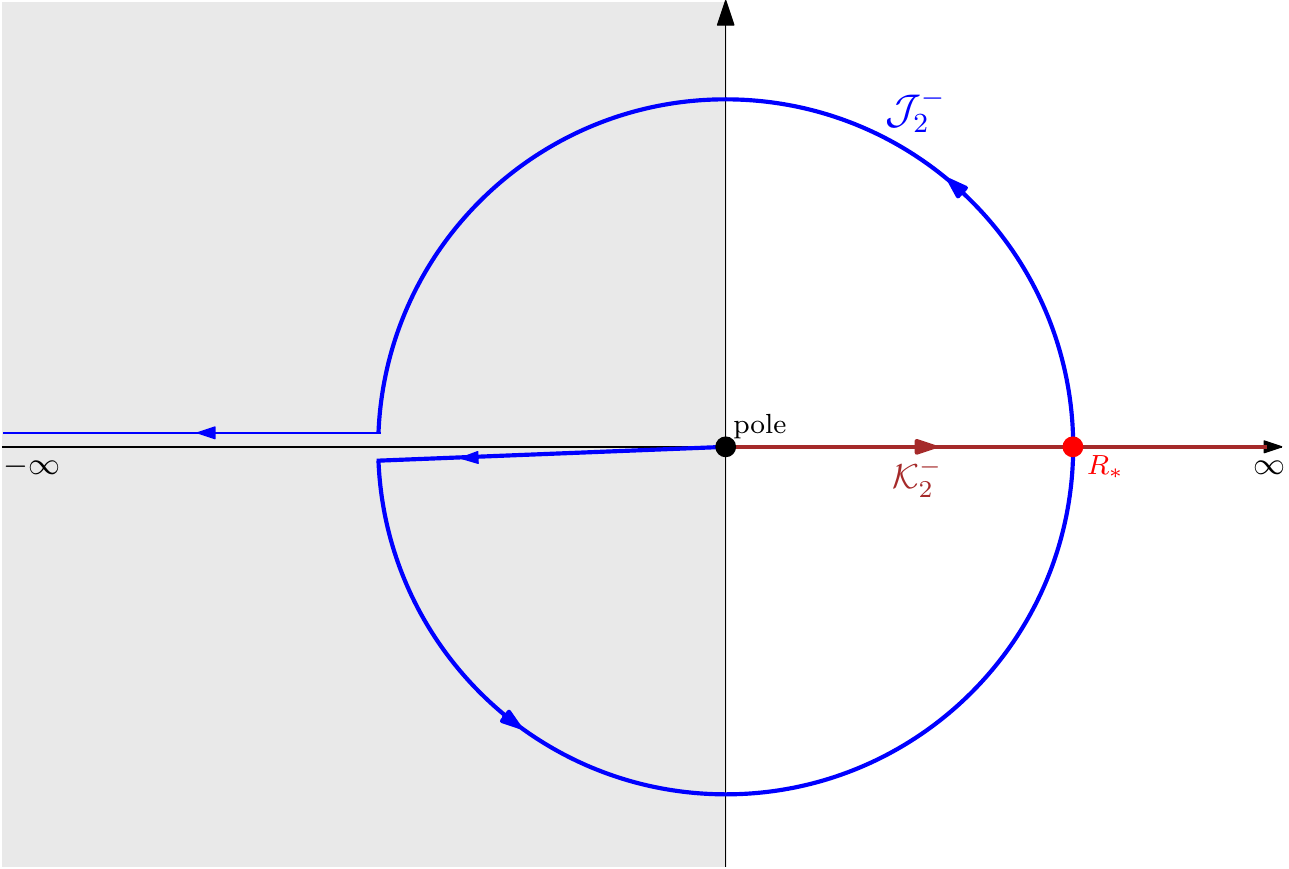}
		\hspace{0.6cm}
		\includegraphics[width=0.47\textwidth]{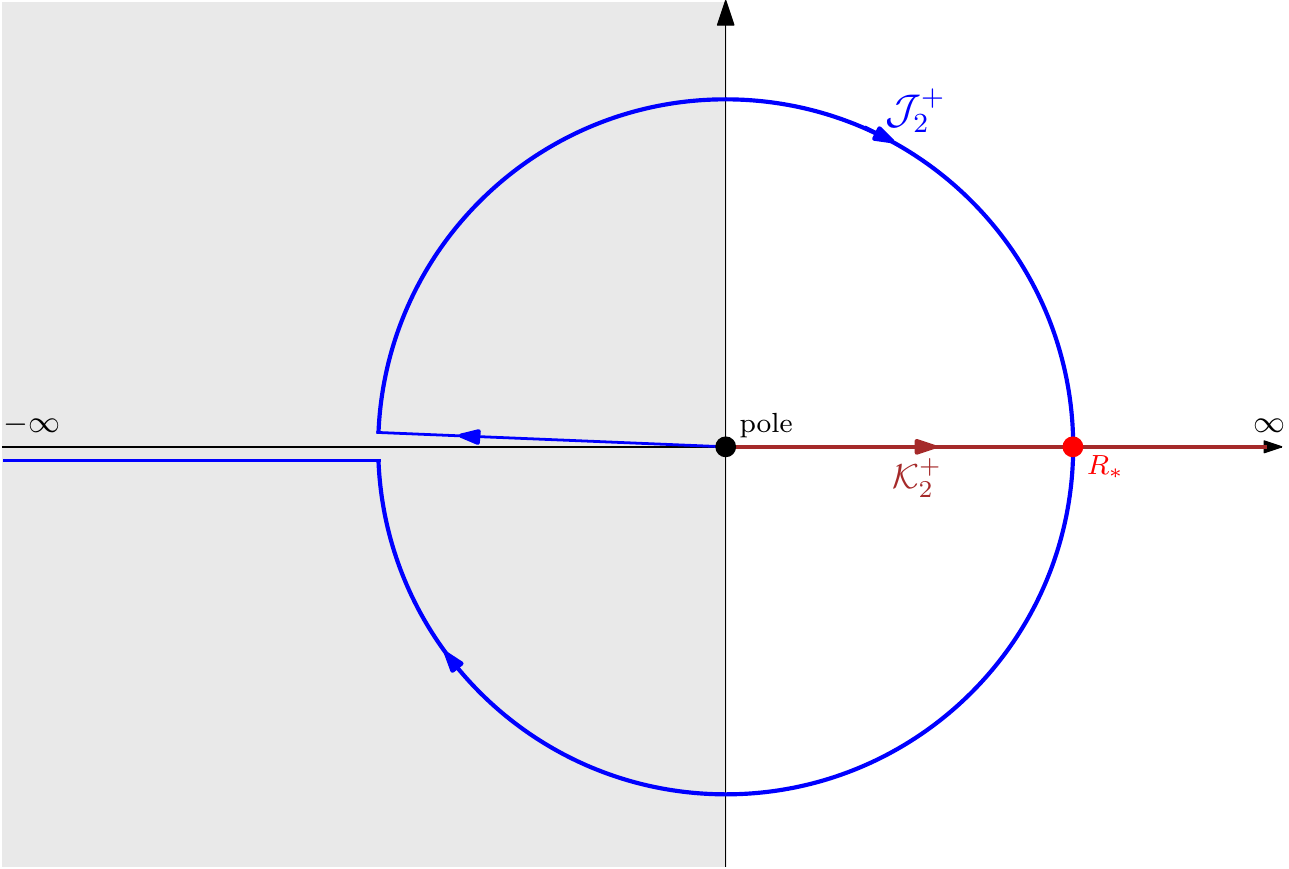}
		\hspace{1cm}
		\caption{ For attractive interactions,  ${\cal J}_2^{\mp}$ are the two types of steepest descent cycles. 
			${\cal J}_2^{-}$ and  ${\cal J}_2^{+}$ only differ by the  direction of the circle. This generates the ambiguity in the $ [{\cal M} \overline 
			{\cal M}]_{\pm} $ event. 
		}
		\label{fig:thimbles-2}
	\end{figure}

	For the ${\cal J}_2^{\mp}$ cycles, the integral is not independent of the ${\cal J}_1 (0)$ cycle. 
	Integration over the cycles shown in Fig.\eqref{fig:thimbles-2} yields
	\begin{align}
		\int_{{\cal J}_2^{\mp}}   dz  z^2  e^{+\frac{1}{g^2} f(z) }  = {\cal V } - I_{\mp}(g^2) 
	\end{align} 
	where the two-fold ambiguity arise from the orientation of the circle segment in ${\cal J}_2^{\mp}$. If we denote circle segment as $C_{\pm}$ depending on its orientation, the pole contribution 
	\begin{align}
	\oint_{C_{\pm}}    dz z^2   e^{ \frac{1}{g^2} \frac{1}{z} }   = \pm 2\pi i   \frac{1}{6}     \left( \frac{1}{g^2} \right)^3    
	 \end{align}
The flip of orientation of this  segment  in ${\cal J}_2^{\mp}$ is the reason for the ambiguity in the $ [ {\cal M}_{\alpha_i} \overline {\cal M}_{\a_{i}}]_{\pm} $ amplitude in \eqref{RadialIntegral_TwoFoldAmbiguity}. 

 The fact that ${\cal J}_2 (\theta )$ changes abruptly as $\theta$ crosses $0$  means that for the attractive interaction, the Stokes line is at 
 ${\rm Arg}(g^2) =0$.  This simple fact means that we are trying to define the theory on a Stokes line.  The phenomena  we encounter  (two-fold ambiguity that arises from non-Borel summability of perturbation theory, and two-fold ambiguities in   the $ [ {\cal M}_{\alpha_i} \overline {\cal M}_{\a_{i}}]_{\pm} $ amplitudes) are natural consequences of an effort to try to define the QFT  on a Stokes line 
 where it is  most subtle.   Yet, it is still possible to give it an unambiguous meaning by left/right Borel-Ecalle resummation.

	\section{3 Instanton Order} \label{Section: 3Instanton}

In this section, we compute the partition function at the 3 instanton level and specifically, terms in the 3-cluster $[\C_3]$.  The partition function is written as
\begin{equation}\label{3instanton}
Z_3 = \frac{[\C_1]^3}{3!}  \int \mrmd \mbfr_i \mrmd \mbfr_j \mrmd \mbfr_k \, e^{- V_{\mathrm{int}} } \quad  , \quad [\C_1] = a_1(g)e^{-S_0} = a_1(g)\x
\end{equation}
where $V_{\mathrm{int}}$ is the  sum of   Coulomb interaction of the monopoles located at $\mbfr_i$, $\mbfr_j$ and $\mbfr_k$.

\paragraph{Repulsive Interactions}We first assume that all the interactions between monopoles are repulsive. Then, the interaction potential is written as
\begin{equation}\label{3InstantonInteraction_SameSign}
V_{\mathrm{int}} = \frac{\pi}{g_3^2}\left[ \frac{|\a_i \a_j|}{|\mbfr_i - \mbfr_j|} + \frac{|\a_j \a_k|}{|\mbfr_k -  \mbfr_j|} + \frac{|\a_i \a_k|}{|\mbfr_k - \mbfr_i|} \right].
\end{equation} 
To simplify the problem, we adapt the Jacobi coordinates as
\begin{equation} \label{JacobiCoordinates}
\mbfq_0 = \frac{1}{3}(\mbfr_i + \mbfr_j + \mbfr_k)\; , \; \mbfq_1 = \mbfr_j - \mbfr_i \; , \; \mbfq_2= \mbfr_k  - \frac{1}{2}(\mbfr_j + \mbfr_i).
\end{equation} 
Then, the spatial integral becomes
\begin{equation}\label{3instanton_transformed}
J_3(g_3^2) = \V \int \mrmd \mbfq_1 \mrmd \mbfq_2 \, \exp\left\{ -\frac{\pi}{g_3^2}\left[\frac{|\a_i\a_j|}{|\mbfq_1|} + \frac{|\a_j \a_k|}{|\mbfq_2 + \frac{1}{2}\mbfq_1|} + \frac{|\a_i \a_k|}{|\mbfq_2 - \frac{1}{2}\mbfq_1|} \right]   \right\}
\end{equation}
The terms $|\mbfq_2 + \frac{1}{2}\mbfq_1| = |\mbfr_k - \mbfr_j|$ and $|\mbfq_2 - \frac{1}{2}\mbfq_1| = |\mbfr_k - \mbfr_i| $ can be expressed in a more convenient way using the geometry in Fig \ref{3instanton_geometry}.
\begin{figure}[h]
	\begin{center}
		\includegraphics[scale=2]{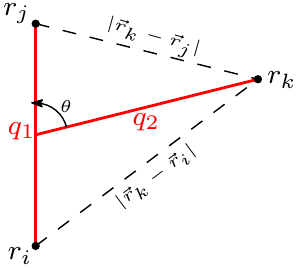}
	\end{center}
	\caption{Three monopoles located at  $\mbfr_i,  \mbfr_j ,  \mbfr_k$. Their relative distances can be expressed by the Jacobi coordinates $\mbfq_1$ and $\mbfq_2$, which simplifies the 3 body problem.}
	\label{3instanton_geometry}
\end{figure}
Using the cosine rule, we write
\begin{align}
|\mbfq_2 + \frac{1}{2}\mbfq_1| &= |\mbfr_k - \mbfr_j| = q_2^2 + \left(\frac{q_1}{2}\right)^2 + q_1 q_2\cos\t \\
|\mbfq_2 - \frac{1}{2}\mbfq_1| &= |\mbfr_k - \mbfr_i| = q_2^2 + \left(\frac{q_1}{2}\right)^2 - q_1 q_2\cos\t 
\end{align}
where we set $q_{1,2} = |\mbfq_{1,2}|$.
Then, last two terms in the exponent in \eqref{3instanton_transformed} becomes
\begin{align}
\frac{|\a_j \a_k|}{|\mbfq_2 + \frac{1}{2}\mbfq_1|} + \frac{|\a_i \a_k|}{|\mbfq_2 - \frac{1}{2}\mbfq_1|} = \frac{1}{q_2} \bigg(|\a_j \a_k|\, F_+(q_1,q_2)+  |\a_i \a_k|\, F_-(q_1,q_2) \bigg)
\end{align}
where
\begin{equation}\label{DipolelikeTerm}
F_\pm(q_1,q_2)  = \left( 1 \pm \frac{q_1}{q_2}\cos\t + \frac{q_1^2}{4q_2^2} \right)^{-1/2} .
\end{equation} 
Since the large separation region dominates the integrals, both $q_1$ and $q_2$ are large and have approximately same size. Then, expanding $F_\pm$ around $q_1 = q_2$ we get 
\begin{equation}\label{DipoleLikeExpansion}
F_\pm(q_1,q_2) = f_\pm(\t) + O\left(\frac{q_1}{q_2}\right)
\end{equation} and at the leading order, up to a $\t$ dependent pre-factor, the spatial integral reduces to
\begin{equation}\label{RadialIntegral_3Instanton}
J_3(g_3^2) \simeq  \V\int_0^\infty \mrmd q_1\, q_1^2 e^{-\frac{2\pi |\a_i\a_j|}{g_3^2\, q_1}} \int_0^\infty \mrmd q_2 \, q_2^2 \, e^{-\frac{2\pi \left(|\a_j\a_k| + |\a_i\a_k|\right) }{g_3^2 \,q_2}}
\end{equation}
These are in the same form with the integrals we get at 2 instanton level in Section \ref{Section: CriticalPointsAtInfinty}. Thus, we can immediately deduce the partition function from \eqref{radialIntegral_cutoffResult3} as
\begin{equation}\label{subext3}
Z_3 = \frac{\xi^3}{3!} \, \V \left(\V + I_1(g^2)\right)\left(\V + I_2(g^2)\right)
\end{equation}
where $I_1$ and $I_2$ are 
\begin{align}
I_1(g^2) &= \frac{4 \pi v^{-3}}{6} 
\left(\frac{2 \pi |\a_i\cdot\a_j|}{g^2}\right)^{3}   \left(  \ln \Big(    \frac{2 \pi |\a_i\cdot\a_j| }{g^2 }     \Big) + \g -   \frac{11}{6}   \right) ,\label{integral1_3cluster} \\
I_2(g^2) &=\frac{4 \pi v^{-3}}{6} 
\left(\frac{2 \pi}{g^2} \Big(|\a_j\a_k| + |\a_i\a_k|\Big)  \right)^{3}    \Bigg(  \ln \bigg[    \frac{2 \pi }{g^2 } \Big(|\a_j\a_k| + |\a_i\a_k|\Big)      \bigg] + \g -   \frac{11}{6}   \Bigg) \label{integral2_3cluster} .
\end{align}

\paragraph{Attractive Interactions:}Now, we turn our attention to the cases with attractive Coulomb potentials. An example suitable to the our choice of Jacobi coordinates in \eqref{JacobiCoordinates} is the case which there are 2 monopoles located at $r_i$ and $r_j$ and 1 anti-monopole at $r_k$. (See Fig. \ref{3instanton_geometry}.) The associated interaction potential is given as
\begin{equation}\label{3InstantonInteraction_DifferentSign}
V_{\mathrm{int}} = \frac{\pi}{g_3^2}\left[ \frac{|\a_i \a_j|}{|\mbfr_i - \mbfr_j|} - \frac{|\a_j \a_k|}{|\mbfr_k -  \mbfr_j|} - \frac{|\a_i \a_k|}{|\mbfr_k - \mbfr_i|} \right]
\end{equation}

The partition function in this case can be probed by an analytical continuation of the $q_2$ integral in \eqref{RadialIntegral_3Instanton} as we did for attractive two instanton cases. This leads to a two fold ambiguous result in the form of \eqref{RadialIntegral_TwoFoldAmbiguity} and we get
\begin{equation}\label{PartitionFunction_3Instanton}
Z_3 = \frac{\xi^3}{3!} \, \V \left(\V + I_1(g^2)\right)\left(\V + I_2(g^2) \pm i\pi \left(\frac{\pi}{g^2} \big(|\a_j\a_k| + |\a_i\a_k|\big) \right)^{3} \right)
\end{equation}

\noindent{\bf Remark:} There are other possible sign combinations for the interaction potential in \eqref{3InstantonInteraction_DifferentSign}. In general, for any 3 instanton anti-instanton event, either all the signs are $(+)$ as in \eqref{3InstantonInteraction_SameSign} or there are 2 $(-)$ sign along with 1 $(+)$ sign as in \eqref{3InstantonInteraction_DifferentSign}. In the latter case, the Jacobi coordinates should be chosen appropriately so that the signs in the exponent of the $q_2$ integral should be the same. Otherwise, it is possible that they could cancel out and higher-order terms arising from \eqref{DipoleLikeExpansion} could lead to $\frac{1}{q^2_2}$ terms, i.e. dipole interaction. In these cases, the logarithmic behavior arising from the incomplete gamma function would be obscured. With appropriate coordinate choices, all of them lead to the same form with \eqref{PartitionFunction_3Instanton} up to modifications in charge-dependent terms.
\paragraph{Relation to the Cluster Expansion:} Now let us understand our result in terms of the cluster expansion: First note that if we consider them separately, both $q_1$ and $q_2$ integrals in \eqref{RadialIntegral_3Instanton} can be interpreted as 2 body interactions. In this sense, the $q_1$ integral, along with the volume prefactor $\V$, corresponds to the interaction between the monopoles at $\mbfr_i$ and $\mbfr_j$. Its contribution is given as $\xi^2\V\big(\V + I_1(g^2)\big) $, which is in parallel with the 2 instanton events in Section \ref{Section: CriticalPointsAtInfinty}. On the other hand, if we took the $q_2$ integral independently, it would correspond to the radial integration of an interaction between the monopole at $\mbfr_k$ and a monopole with a total charge $Q_{ij} =\frac{2\pi}{g_3}\left(\a_i + \a_j\right)$. In \eqref{RadialIntegral_3Instanton}, this means that the monopole at $\mbfr_k$ coalesce into the 2 body interaction of the monopoles at $\mbfr_i$ and $\mbfr_j$. 

This is indeed how we should relate \eqref{RadialIntegral_3Instanton} to the cluster expansion. Specifically, the contribution of the $q_1$ integral consists of 2-cluster connected and disconnected parts, i.e. $\xi^2 \V\big(\V + I_1(g^2)\big) \sim \V^2 [\C_1]^2 + \V [\C_2] $, as in 2 instanton case. The contribution of the $q_2$, i.e. $\xi(\V + I_2(g^2))$, on the other hand, should be interpreted in terms of the connectedness of the monopole at $\mbfr_k$ to the two body interaction represented by the $q_1$ integral. Then,
\begin{itemize}
	\item Volume contribution of $q_2$ integral refers that the monopole at $\mbfr_k$ is not connected to the neither of the other monopoles. Thus, it contributes as an independent 1-cluster, i.e. $\V[\C_1]$, onto the expansion of the $q_2$ integral, i.e. $\V^2 [\C_1]^2 + \V [\C_2]$. 
	
	\item $I_2(g^2)$ part refers to the cases that the monopole at $r_k$ is attached to one or both of the other monopoles. Therefore, when $\xi I_2(g^2)$ combines with the two 1-cluster part $\xi^2 \V^2$, it results in a product of 1-cluster and 2-cluster, i.e. $\V^2 [\C_1]^2  \rightarrow \V^2 [\C_1][\C_2]$. Similarly, the combination of  $\xi I_2(g^2)$ with $\xi^2 \V I_1(g^2)$ turns the 2-cluster part of the $q_1$ integral into a 3-cluster, i.e. $\V [\C_2] \rightarrow \V[\C_3]$. 
\end{itemize}
This indicates that the 3 instanton partition function in \eqref{subext3} indeed takes the form of $\xi^3$ order terms in the cluster expansion \eqref{EFT2}:
\begin{equation}\label{3Instanton_ClusterForm}
Z_3 = \frac{\V^3}{3!} [\C_1]^3 + \frac{\V^2}{2!} [\C_1][\C_2] + \frac{\V}{1!} [\C_3] 
\end{equation}
with the following identifications:
\begin{align}
\xi^3\big(I_1(g^2) + I_2(g^2) \big)&\in [\C_1][\C_2], \\
\xi^3 I_1(g^2) I_2(g^2) 	&\in [\C_3] .
\end{align}

It is also possible to express \eqref{3Instanton_ClusterForm} as a collection of the monopole events. First note that as stated in the footnote in page \pageref{footnote: BPS_limit}, the effective interaction between $\mcalM_{\a_i}$ and $\mcalM_{\a_j}$ vanishes. Therefore, they do not form 2 clusters which are restricted as in \eqref{clustering2}. However, contrary to the 2-cluster case, $\mcalM_{\a_i}$ and $\mcalM_{\a_j}$ can participate in the same 3-cluster term. 
Then, we get the following characteristics:
\begin{itemize}
	\item Terms arising from the combination of 1-cluster events, e.g. $\frac{\V^3}{3!}[\mcalM_{\a_i}]^3$
	\item Unambiguous 2-cluster terms with an additional 1-cluster term, e.g
	\begin{equation}
	\V^3[\mcalM_{\a_j}][\mcalM_{\a_{i+1}}][\mcalMbar_{\a_i}] + \V^2[\mcalM_{\a_j}][\mcalM_{\a_{i+1}}\mcalMbar_{\a_i}] \quad , \quad j\neq i\pm 1
	\end{equation}
	which is a direct extension of the second order term.
	\item Ambiguous 2-cluster terms with an additional 1-cluster term, e.g
	\begin{equation}
	\V^3[\mcalM_{\a_j}][\mcalM_{\a_{i}}][\mcalMbar_{\a_i}] + \V^2[\mcalM_{\a_j}][\mcalM_{\a_{i}}\mcalMbar_{\a_i}]_\pm \quad , \quad j\neq i\pm 1
	\end{equation}
	which is a direct extension of the second order term. 
	\item Unambiguous 3-cluster terms, e.g.
	\begin{equation}
	\V^3[\mcalM_{\a_{i+1}}][\mcalM_{\a_{i+1}}][\mcalMbar_{\a_i}] + \V^2[\mcalM_{\a_{i+1}}][\mcalM_{\a_{i+1}}\mcalMbar_{\a_i}] + \V [\mcalM_{\a_{i+1}}\mcalM_{\a_{i+1}}\mcalMbar_{\a_i} ]
	\end{equation}
	\item Ambiguous 3-cluster terms, e.g.
	\begin{equation}
	\V^3[\mcalM_{\a_{i}}][\mcalMbar_{\a_i}[\mcalM_{\a_{i}}]] + \V^2[\mcalM_{\a_{i}}][\mcalMbar_{\a_i}\mcalM_{\a_{i}}]_\pm + \V [\mcalM_{\a_{i}}\mcalMbar_{\a_i}\mcalM_{\a_{i}} ]_\pm
	\end{equation}
\end{itemize}

Since only $3$-cluster terms $[\C_3]$ contribute to the spectrum at 3 instanton order, the ambiguous imaginary part of the spectrum arises from $\xi^3 I_1(g^2) I_2(g^2) $ part of the partition function. Then, considering events with the ambiguity, we get the imaginary contribution to spectrum as
\begin{align}
\Im  \mcalE^{(3)} &=  \Im[\mcalM\mcalMbar\mcalM]_\pm 
 \simeq \pm \left(\frac{\mrms_0}{g^2} \right)^{12} \ln \left(\frac{1}{g^2}  \right) \, e^{-\frac{12\pi}{g^2}}. \label{3Instanton_SpectrumAmbiguity}	
\end{align}

    	\bibliographystyle{JHEP}
   	\bibliography{QFT-Mithat}
\end{document}